\documentclass[fleqn,usenatbib]{mnras}

\usepackage{newtxtext,newtxmath}
\usepackage[T1]{fontenc}

\DeclareRobustCommand{\VAN}[3]{#2}
\let\VANthebibliography\thebibliography
\def\thebibliography{\DeclareRobustCommand{\VAN}[3]{##3}\VANthebibliography}

\usepackage{graphicx}	% Including figure files
\usepackage{amsmath}	% Advanced maths commands

\title[Bipolar outflow in NGC\,1125]{A jet-driven bipolar outflow in NGC\,1125} 

\author[Schönell et al.]{
Astor J. Schönell Jr,$^{1,2}$\thanks{E-mail: astor.schonell@gmail.com}
Rogemar A. Riffel,$^{2}$\thanks{E-mail: rogemar@ufsm.br}
Rogério Riffel,$^{3}$
Thaisa Storchi-Bergmann$^{3}$
\\
$^{1}$Instituto Federal de Educação, Ciência e tecnologia Farroupilha (IFFar) \\
$^{2}$Universidade Federal de Santa Maria (UFSM)\\
$^{3}$Universidade Federal do Rio Grande do Sul (UFRGS)\\
}
\date{Accepted XXX. Received YYY; in original form ZZZ}

\pubyear{2024}

\begin{document}
\label{firstpage}
\pagerange{\pageref{firstpage}--\pageref{lastpage}}
\maketitle

\begin{abstract}

To study the role of the feedback from the Active Galactic Nuclei (AGNs) in the evolution of its host galaxy, we need observational constraints on 100 pc scales. We used the Gemini Near Infrared Integral Field Spectrograph in the J and K bands at a spatial resolution of 100 pc and spectral resolution of 45 km\,s$^{-1}$ to observe the central region of the Seyfert galaxy NGC1125. Emission-line flux distributions in ionized and molecular gas extends up to $\approx$ 300\,pc from the nucleus, where they are found to peak. The Pa$\beta$ and [Fe\,{\sc ii}]$\lambda$1.2570$\mu$m emission-lines show two components: a narrow and a broad. The narrow component is preferably extended from the north-east to the south-west, while the broad component is perpendicular to it. Their kinematics are also different, with the narrow component showing a rotation pattern, with low velocity dispersion values ($\sigma$ $\approx$ 140 km s$^{-1}$) and the broad component a disturbed velocity field and high values of $\sigma$ ($\approx$ 250 km s$^{-1}$). We interpreted the narrow component velocity fields as due to gas rotating in the galaxy plane and fitted rotation velocity models to it, plus an outflow component in the ionized gas. The broad component is interpreted as an outflow, with mass outflow rate in the range of 0.6 to 1.1 M$_{\sun}$ yr$^{-1}$, with an outflow power ranging from 3.9$\times$10$^{40}$ to 1.1$\times$10$^{41}$ erg\,s$^{-1}$, which represents 0.07\% and 0.2\% of the bolometric luminosity of the AGN. There is an explicit relation between the shock ionized outflow and the low-luminosity radio source. 
\end{abstract}

\begin{keywords}
galaxies: active -- galaxies: evolution -- galaxies: Seyfert -- galaxies: nuclei
\end{keywords}

%%%%%%%%%%%%%%%%%%%%%%%%%%%%%%%%%%%%%%%%%%%%%%%%%%

%%%%%%%%%%%%%%%%% BODY OF PAPER %%%%%%%%%%%%%%%%%%
\section{Introduction}
\label{sec-intro}

It is well established that the evolution of an Active Galactic Nucleus (AGN) host galaxy can be affected by the radiation, jets and outflows provided by the AGN in a process called feedback \cite[e.g.][]{fabian2012,silk2012,harrison2024}. The AGN feedback impact, can be positive \citep[e.g.][]{maiolino2017,gallagher2019}: when the winds compress the gas triggering star formation and thus increasing the star formation rate (SFR) or, more commonly, negative \citep[e.g.][]{cicone2012,cano2012,harrison2017}: when the AGN radiation, jets or winds maintain the gas heated or remove the gas supply from the host galaxy, thus decreasing the SFR or even extinguishing the star formation. In parallel, there is the AGN feeding process via gas accretion, that triggers the nuclear activity. These two processes -- feeding and feedback -- can help to explain the correlation between the mass of the super massive black hole (SMBH) and the mass of the galaxy bulge  \citep{ferrarese2005,somerville2008,kormendy2013}. 

An effective method to observe and quantify the feeding and feedback processes is by using near-infrared (near-IR) integral field spectroscopy (IFS) of nearby galaxies. From such observations, one can map the ionized and molecular gas distributions, excitation and kinematics in the vicinity of the AGN ($\approx$  100 pc scales) providing important constrains on the physics of the feeding and feedback processes. 
%To reach two dimensional coverage with spectral and spatial resolutions to resolve gas inflows and mass outflows in scales of a few to tens of parsecs in nearby galaxies \citep{sanchez2009,davies2009,davies2014}, it is used 8-10 meters telescopes IFS with adaptive optics (AO). 
The near-IR spectral region has the advantage of being less affected by dust extinction -- that is usually high in central region of galaxies -- than optical observations (most present in the literature, i.e. CALIFA, SAMI, MANGA and DIVING3D, for example), allowing to resolve the central region down to a few parsecs and simultaneously map the gas in two distinct phases: ionized and molecular (H$_2$), where the latter is observable in the near-IR K band but is not in the optical, for example.

In AGN host galaxies, the near-IR emission at hundreds of parsecs scales is originated by the heating and ionization of ambient gas by the AGN radiation and by shocks \citep{riffel2006,riffel2010,ardila2005,rogerio2006,rogerio2013}. In recent studies \citep[e.g.][]{riffel2018,schonell2019} the observations have been showing distinct spatial distributions between the ionized and molecular gases at these scales.

In  general, the kinematics of the molecular gas is dominated by rotation in the disks, but can also present inflows and outflows, while the ionized gas traces a more disturbed medium usually associated with outflows from the AGN, nonetheless presents also a disk rotation component \citep[e.g.][]{riffel2010,riffel2013,barbosa2014,mazzalay2014,diniz2015,reunanen2002}. The molecular emission is manly due to thermal processes -- X-ray heating and shocks -- and its distribution is more extended than the ionized gas.  In \cite{riffel2023}, we used a sample of 33 X-ray selected AGN of the local universe, observed with the Gemini near-Infrared integral field spectrograph (NIFS), to analyze the molecular and ionized gas kinematics using non-parametric measurements of the H$_2$2.1218$\mu$m and Br$\gamma$ emission-lines, to spot locations where the gas is strongly impacted by outflows for both gas phases. We identified kinetically disturbed regions (KDR) in 31 galaxies for the ionized gas (94 per cent of the sample) and 25 galaxies with KDRs for the molecular gas (76 per cent of the sample). We attributed the KDR  as being produced by AGN outflows, where the mass outflow rates and power presented positive correlations with the AGN bolometric luminosity.

NGC\,1125 is part of a volume complete sample of 20 nearby (z $\le$ 0.015) Seyfert galaxies drawn from the Swift-BAT 60-month catalog with 14-195\,keV luminosities larger than 10$^{41}$ erg s$^{-1}$, observed in the J and K bands with the goal of mapping the ionized and hot molecular gas distributions and kinematics. A detailed description of this sample is presented in  \citet{riffel2018}. All these 20 galaxies are also in the sample of 33 galaxies mentioned above. The additional 13 galaxies are objects included in the 105-month catalog of the Swift Burst Alert Telescope (BAT) survey \citep{oh2018} at redshifts z$\le$0.12, with NIFS K-band data available in the Gemini Science Archive.
Previous works using this sample include the analysis of the stellar kinematics \citep{riffel2017}, stellar population properties \citep{rogerio22}, hot molecular gas emission origin \citep{riffel21b}, and general properties of the hot molecular and gas kinematics \citep{riffel2023} in the inner few hundreds of parsecs of these galaxies. Additionally, individual galaxy studies have provided insights into the detailed circumnuclear physics of both molecular and ionized gas \citep{riffel2014,diniz2015,ardila2016,schonell2017,dahmer2019}. In this work, we conduct an in-depth analysis of NGC\,1125, a galaxy exhibiting clear signatures of bipolar outflows in ionized gas. Detailed studies of specific galaxies like NGC\,1125 are crucial as they enhance our understanding of how outflows interact with the interstellar medium, their impact on star formation processes and for regulating galactic growth.
%Within these parameters for the observations, we pretend to answer in this paper, particularly for NGC\,1125, questions like: how do the outflows interact with the interstellar medium? What are the mechanisms behind the excitation of the observed emission-flux lines?  What are the mass outflow rates and kinetic power? How are the measured properties related with the luminosity of the AGN? Can the outflows influence in the galaxy evolution? By answering these questions in a detailed and thorough study of the individual object NGC\,1125, we can further expand the analysis made in the above-mentioned paper \citet{riffel2023} that used only the K band to explore the origin of the molecular gas on several AGN.
With these goals in mind, we present the gaseous distribution, excitation and kinematics of the inner $\approx$ 300\,pc radius of the Seyfert 2 galaxy NGC\,1125 (MCG -03-08-035), a spiral galaxy (SB0/a) at a distance of 47.1 Mpc, where 1.0 arcsec corresponds to 228 parsecs at the galaxy \citep{theureau98}. It has a radio source with a linear structure at a position angle (PA) 120$^{\circ}$ \citep{thean2000} which does not correspond to the the slightly resolved [O\,{\sc iii}] emission at a PA of 56$^{\circ}$ \citep{mulchaey1996} with H${\alpha}$ emission extended along the disk of the galaxy (PA = 48$^{\circ}$), with a hard X-Ray luminosity of 4.37$\times$10$^{42}$ erg s$^{-1}$. 

%It presents a complex star formation history \citep{rogerio22}, within 100\,pc from the nucleus towards the north-west being dominated by a young (t $\le$ 50 Myr) stellar population component (SPC). A ring-like structure is observed in the intermediate-age stars (100Myr $\le$ t $\le$ 2 Gyr) which dominate the light in the central region of the galaxy. Older (t $\ge$ 5 Gyr) stellar populations are also observed in regions from the centre to the north. These are the overall results for the stellar populations of NGC\,1125, which we will not discuss here, since the main goal of this work is the study of the gas kinematics and excitation.

This paper is organized as follows: in Section 2, we describe the observations and data reduction procedures. In section 3 we discuss how the emission-line fitting was done. The results are presented in Section 4 and discussed in Section 5. We present our conclusions in Section 6. We use a $h=0.7$, $\Omega_{\rm m}=0.3$ and $\Omega_{\Lambda}=0.7$ cosmology throughout this paper.

\section{Observations and data reduction}

The observations of NGC\,1125 were obtained with the Gemini NIFS \citep{mcgregor2003} integral field spectrograph. It has a square field of view of $\approx$ 3.0 $\times$ 3.0 arcsec$^2$, divided into 29 slices with an angular sampling of 0.103 $\times$ 0.042 arcsec$^2$. The observations of NGC\,1125 were done in August 12 and 31, 2018 for the J and K band respectively, in the queue mode under the project GN-2018B-Q-140. For the J band, we used the J\_G5603 grating and ZJ\_G0601 filter while for the K band we used the Kl\_G5607 grating and HK\_G0603 filter.

The observations followed the standard object–sky–object dither sequence, with off-source sky positions at 1 arcmin from the galaxy, and individual exposure times of 450\,s. The J-band spectra are centered at 1.25 $\mu$m, covering the spectral range from 1.14 to 1.36 $\mu$m. The K-band data are centered at 2.20 $\mu$m and cover the 2.00–2.40 $\mu$m spectral region. The total exposure time at each band was 60 min.

The data reduction was accomplished using tasks contained in the NIFS package, which is part of GEMINI {\sc IRAF} package, as well as generic {\sc IRAF} tasks \citep{tody86,tody93}. The data reduction followed the standard procedure, which includes the trimming of the images, flat-fielding, cosmic ray rejection, sky subtraction, wavelength, and s-distortion calibrations. In order to remove telluric absorptions from the galaxy spectra, we observed the telluric standard star just after the J-band and K-band observations. The galaxy spectra were divided by the normalized spectrum of the telluric standard star using the NFTELLURIC task of the NIFS.GEMINI.IRAF package. The galaxy spectra were flux calibrated by interpolating a blackbody function to the spectrum of the telluric standard star. We added an additional step in the flux calibration, where we compared and corrected our J and K data using the IR cross-dispersed (XD) long-slit data from \citet{rogerio22}. The J- and K-band data cubes were constructed with an angular sampling of 0.05 $\times$ 0.05 arcsec for each individual exposure. The individual data cubes were combined using a sigma clipping algorithm in order to eliminate bad pixels and remaining cosmic rays by mosaicing the dithered spatial positions, using the peak of the continuum emission as reference.

The angular resolution of the K-band data cube is 0.44 arcsec ($\approx$ 100 pc at the galaxy), as measured from the flux of the telluric standard star. For the J-band emission, the final data cube has also an angular resolution of 0.44$\arcsec$ ($\approx$ 100 pc at the galaxy) and the resulting velocity resolution is 40 km s$^{-1}$ in the J band and 45 $\pm$ km s$^{-1}$ in the K band, measured from the FWHM of typical emission-lines of the Ar and ArXe lamp spectra used to wavelength calibrate the J- and K-band spectra, respectively.

\begin{figure*}
	\centering
	\includegraphics[width=1.0\linewidth]{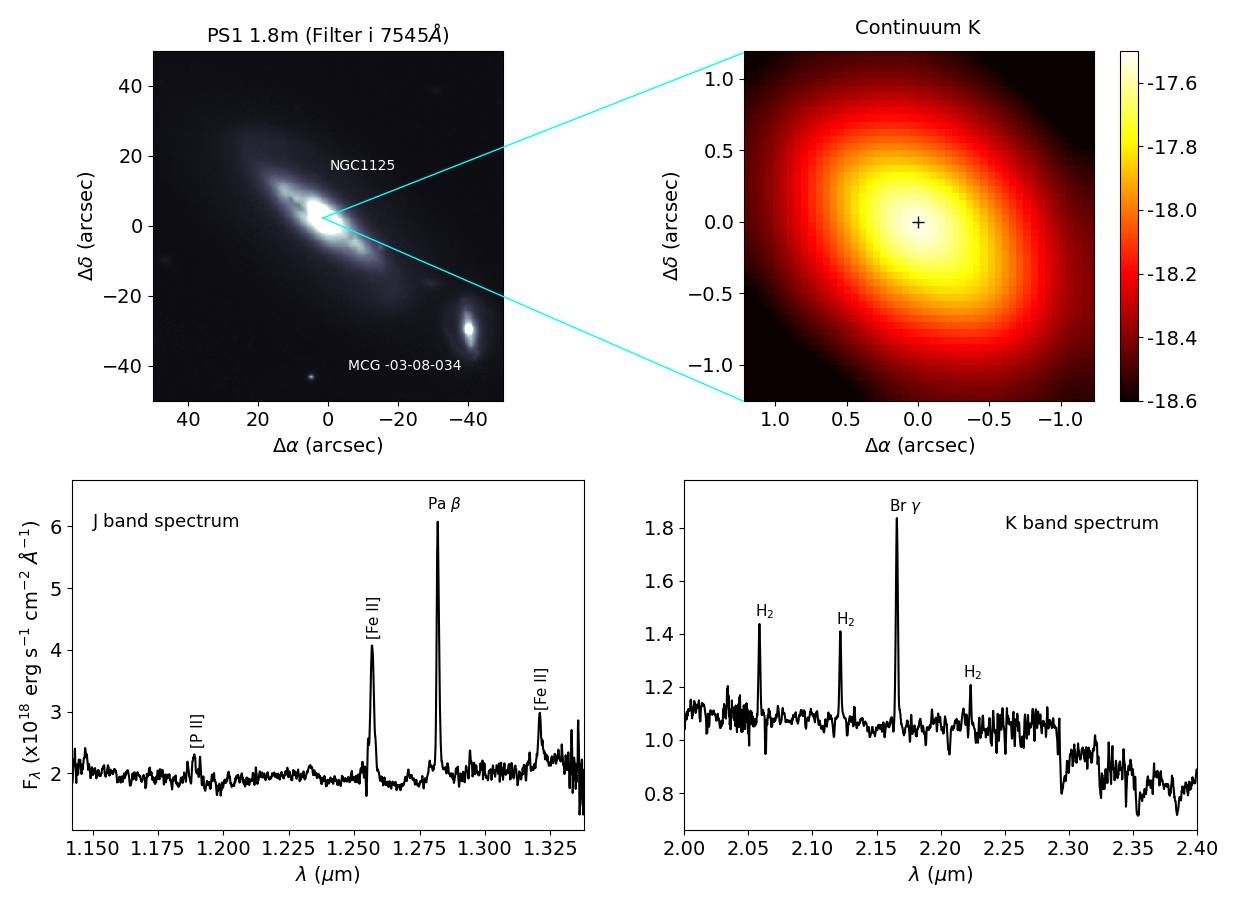}
	\caption{Top-left panel: i image of NGC\,1125 from Pan-STARRS data archive \citep{chambers2016,flewelling2016}. Top-right: NIFS K-band continuum image. The color bar shows the fluxes in logarithmic units of erg s$^{-1}$ cm$^{-2}$ \AA$^{-1}$ spaxel$^{-1}$. Bottom panels: J- and K-band spectra of NGC\,1125 centered at the peak of the continuum and extracted within apertures of 0.18x0.18\,arcsec (40\,pc x 40\,pc at the galaxy).}
	\label{fig1}
\end{figure*}

The top-left panel of Fig.\,\ref{fig1} presents a large-scale image of NGC 1125 obtained from the Pan-STARRS data archive \citep{chambers2016,flewelling2016} and the top-right panel shows a K-band continuum image obtained by computing the average flux of the NIFS data cube at each spaxel. The bottom panels show the J- and K-band spectra in the peak of the continuum emission. The main emission-lines are identified: [P\,{\sc ii}]$\lambda$1.1886 $\mu$m, [Fe\,{\sc ii}]$\lambda$1.2570 $\mu$m, Pa$\beta$ $\lambda$1.2822 $\mu$m, [Fe\,{\sc ii}]$\lambda$1.3209 $\mu$m, H$_2$ $\lambda$2.0338 $\mu$m, He {\sc i}$\lambda$2.0587 $\mu$m, H$_2$ $\lambda$2.1218 $\mu$m, Br$\gamma$ $\lambda$2.16612 $\mu$m and H2$\lambda$2.2233 $\mu$m.

\begin{figure*}
	\centering
	\includegraphics[width=1.0\linewidth]{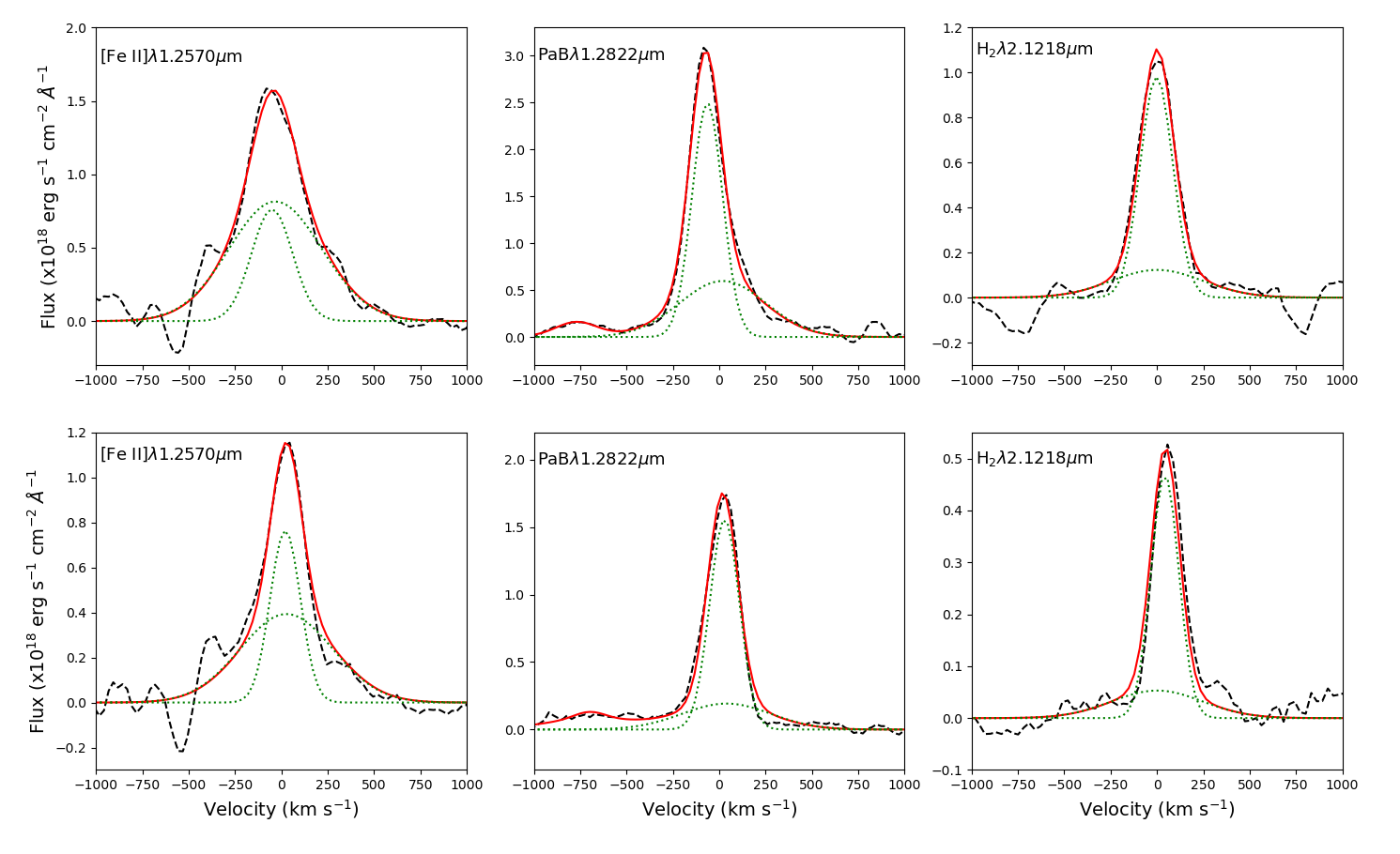}
	\caption{Examples of the fits of [Fe\,{\sc ii}]$\lambda$1.2570 $\mu$m (first row), Pa$\beta$ $\lambda$1.2822 $\mu$m (second row), and H$_2$ $\lambda$2.1218 $\mu$m (third row) emission-line profiles. The position of the top panels are centered at the continuum peak and those of the bottom panels at 0.5$\arcsec$ south-west of the continuum peak. The continuum-subtracted observed profiles are shown as dashed black lines, the fits are in red, and the individual Gaussian components are shown as green dotted lines.}
	\label{fig2}
\end{figure*}

\section{Emission-line fitting}

We used the the {\sc IFSCUBE} package \citep{rd2021} to fit the observed emission-line profiles with Gaussian curves. The fit is performed with two Gaussian functions (for a narrow and a broad component of the NLR) that are necessary to reproduce the observed line profiles at all positions.
First we fitted the emission-line profiles by Gauss-Hermite series. By visual inspecting the resulting maps for the $h_3$ and $h_4$ moments, which trace asymmetric and symmetric deviations from a Gaussian profile, and the individual spectra in distinct locations of the field of view, we identify the need of two Gaussian components to properly represent the emission line profiles. 
The emission-lines in the J- and K bands are fitted separately and all parameters are kept free, as we can see in Fig.\,\ref{fig2} that the line profiles of distinct species clearly show distinct kinematic components.  We used the {\sc CUBEFIT} routine from the {\sc IFSCUBE} package to fit the following emission-lines simultaneously: [P\,{\sc ii}]$\lambda$1.1886 $\mu$m, [Fe\,{\sc ii}]$\lambda$1.2570 $\mu$m, Pa$\beta$$\lambda$1.2822 $\mu$m, [Fe\,{\sc ii}]$\lambda$1.3209 $\mu$m; and then: H$_2$$\lambda$2.0338 $\mu$m, He{\sc i}$\lambda$2.0587 $\mu$m, H$_2$$\lambda$2.1218 $\mu$m, Br$\gamma$$\lambda$2.1661 $\mu$m, and H2$\lambda$2.2233 $\mu$m. Initial guesses of the Gaussian amplitude, centroid velocity, and velocity dispersion are provided to the code based on measurements of the nuclear spectra using the splot IRAF task. After a successful fit of
the nuclear emission-line profiles, the routine fits the surrounding spaxels following a radial spiral loop, using as initial guesses the best-fitting parameters obtained from successful fits of spaxels at distances smaller than 0.3 arcsec from the fitted spaxel, as defined by using the refit parameter. Using this parameter, the fitting routine is optimized to select the most suitable number of components and initial guesses for fitting a specific line profile at a given spaxel. For instance, if one of the Gaussian components has an amplitude close to zero in most neighboring spaxels, the optimal fit will generally converge to a single Gaussian component. This is consistent with the absence of discontinuities in the two-dimensional maps of emission line properties, confirming the reliability of the fits, as smooth distributions in flux and kinematics are expected, at least on scales comparable to the spatial resolution of the data. This indicates that the number of components used is appropriate, so that we can interpret each component by their physical origin. To account for possible continuum emission, we also include a fourth-order polynomial, which is used to fit the continuum before the fitting of the emission-lines.

In Fig.\,\ref{fig2} we show examples of the fits of the [Fe\,{\sc ii}]$\lambda$1.2570 $\mu$m, Pa$\beta$ $\lambda$1.2822 $\mu$m and H$_2$ $\lambda$2.1218 $\mu$m for two different positions: in the top panels are the fits at the location of the peak of the continuum and in the bottom panels are the fits for a position at 0.5$\arcsec$ to south-west of the peak of the continuum. As can be seen, the line profiles are well reproduced by the adopted model. 

\section{Results}
\label{results}

In this section, we present the two-dimensional maps produced using the methodology described in the previous section. In all maps, the grey regions correspond to masked locations where the amplitude of the corresponding emission-line is smaller than 3 times the standard deviation of the continuum next to the line. The north is to the top and east is to the left in all maps. The systemic velocity of the galaxy is subtracted in all velocity maps, assumed to be the value derived from the fitting of the H$_2$$\lambda$2.1218$\mu$m velocity model as described in Sec. 5.1. The velocity dispersions measured for the absorption and emission lines have been corrected for the instrumental width. The emission-lines were fitted with two Gaussian functions (as described in Sec. 3): one corresponding to the narrow component of the emission-line profile - that we have attributed to the galaxy disk emission plus an additional outflow component - and the other corresponding to the broad component - that we have attributed mainly to outflow.

\subsection{Stellar kinematics}

We used the penalized pixel-fitting (PPXF) method of \citet{cappellari2004} to fit the CO absorption band-heads (around 2.3$\mu$m in the K-band spectrum of Fig.\,\ref{fig1}) in order to obtain the line-of-sight velocity distributions of the stars, using the Gemini library of late spectral type stars observed with the Gemini Near-Infrared Spectrograph (GNIRS) Integral Field Unit (IFU) and NIFS \citep{winge2009} as templates.
The stellar line-of-sight velocity distribution was approximated by a Gaussian distribution. PPXF outputs the stellar radial velocity (V$_{*}$), the corresponding velocity dispersion ($\sigma_{*}$), as well as the uncertainties for both parameters at each spaxel. 

\begin{figure*}
	\centering
	\includegraphics[width=1.0\linewidth]{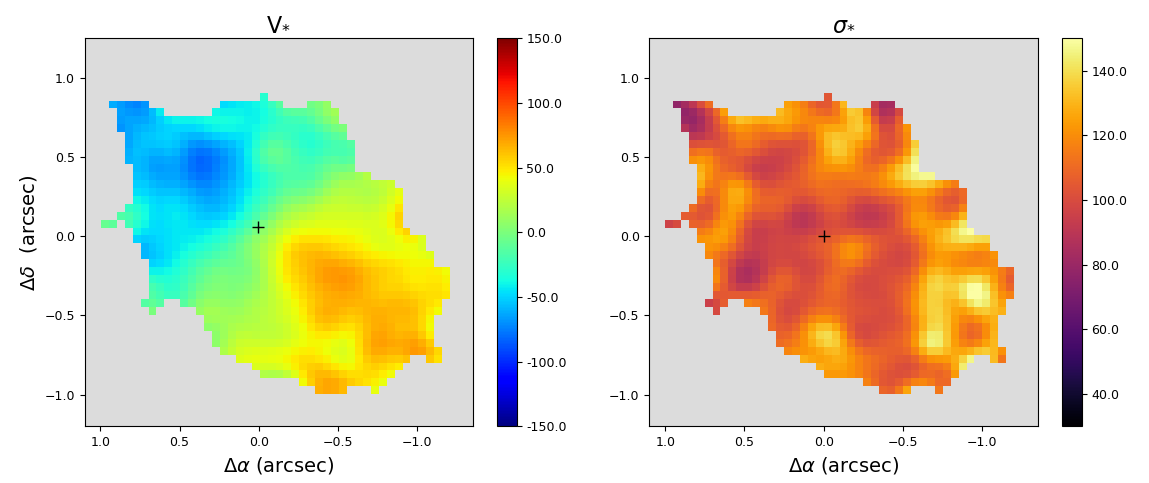}
	\caption{Stellar velocity (V$_{*}$) field (left) and corresponding velocity dispersion ($\sigma_{*}$) map (right). The central cross marks the position of the nucleus, the color bars show V$_{*}$ and $\sigma_{*}$ values in units of km s$^{-1}$ and the grey regions represent the locations where we could not get good fits of the galaxy spectra.}
	\label{stel-kin}
\end{figure*}

In Fig.\,\ref{stel-kin} we present the resulting maps for the V$_{*}$ and $\sigma_{*}$. The grey regions in the maps correspond to masked locations where the uncertainties in V$_{*}$ and $\sigma_{*}$ are larger than 30 km\,$^{-1}$. We have subtracted the systemic velocity of the galaxy, of 3277 km\:s$^{-1}$, obtained from \citet{theureau98}.
The stellar velocity field shows a velocity amplitude of about 100 km\,s$^{-1}$, with red shifts to the south-west and blue shifts to the north-east. The $\sigma_{*}$ map shows values ranging from 60 to 150 km s$^{-1}$ with a median value of $\sigma_{*}$ = 112\,km\,s$^{-1}$ and there is a partial ring of lower $\sigma_{*}$ values (60–80 km\,s$^{-1}$) with a radius of $\approx$ 100 pc surrounding the nucleus, probably due to a ring structure of intermediate-age stars, which dominates the light in the central region of this galaxy \citep{rogerio22}.

\subsection{The gas disk component}

\begin{figure*}
	\centering
	\includegraphics[width=1.0\linewidth]{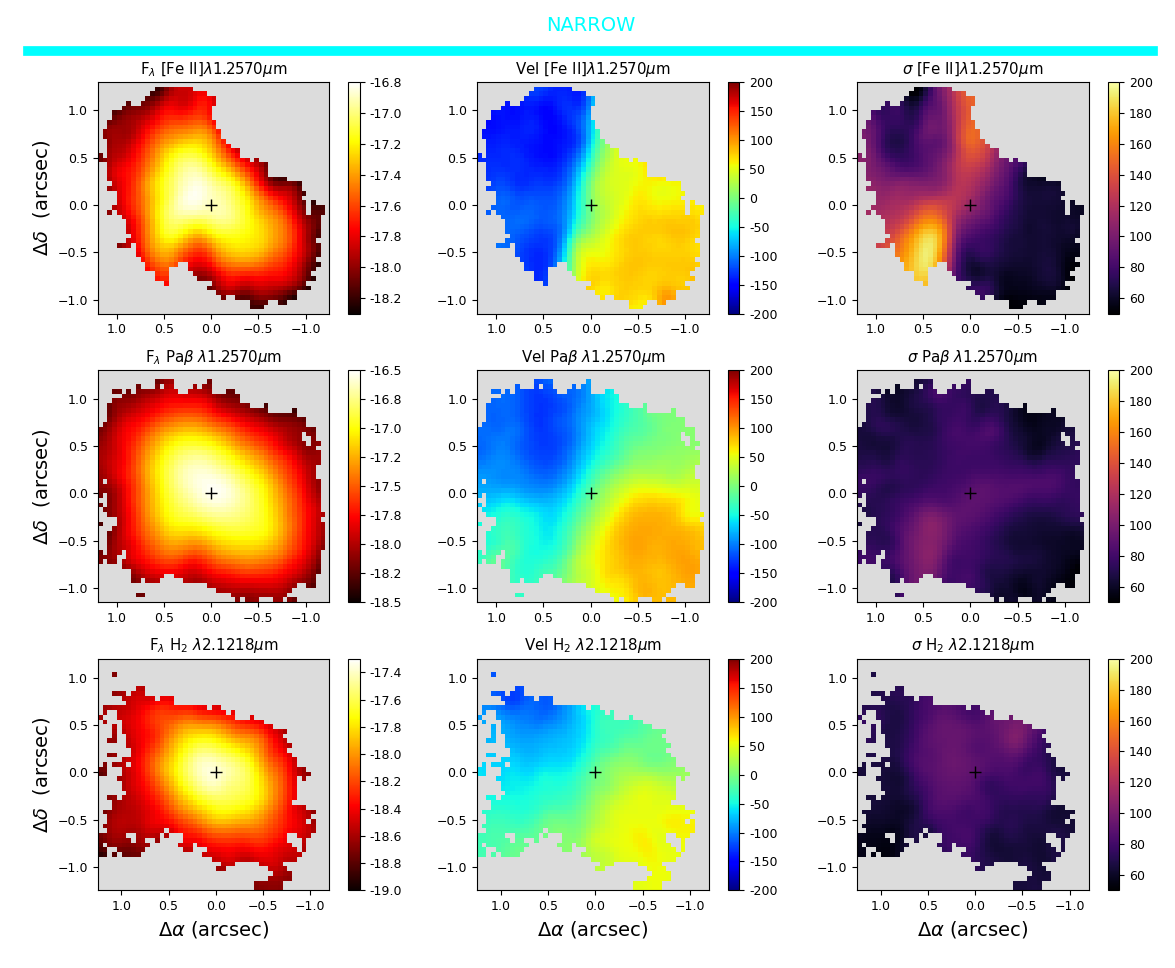}
	\caption{First column, from top to bottom : flux distribution for the [Fe\,{\sc ii}]$\lambda$1.2570$\mu$m, Pa$\beta$ and H$_2$$\lambda$2.1218$\mu$m narrow component, respectively. The color bar shows the fluxes in logarithmic scale in units of erg s$^{-1}$ spaxel$^{-1}$. Second column: velocity maps for the same lines of the first column. The color bar shows the velocity in units of km s$^{-1}$. Third column: $\sigma$ maps for the same emission-lines of the first column. The color bar shows the $\sigma$ values in units of km s$^{-1}$. The central crosses marks the location of the peak of the continuum emission, the grey regions represent masked locations, where the amplitude of the corresponding line profile is smaller than 3 times the noise of the adjacent continuum.}
	\label{fig-flux-n}
\end{figure*}

In the first column of Fig.\,\ref{fig-flux-n} we present the flux distributions for the narrow component (disk dominated component plus a contribution of the outflow) of [Fe\,{\sc ii}]$\lambda$1.2570$\mu$m, Pa$\beta$ and H$_2$$\lambda$2.1218$\mu$m. All maps are most extended along the preferential position angle of $\approx$ 50$^{\circ}$, reaching up to 1$\farcs$5 ($\approx$ 345\,pc) in the case of [Fe\,{\sc ii}] and Pa$\beta$, although being also more extended along PA $\approx$50$^{\circ}$, shows more emission than [Fe\,{\sc ii}] in other directions, being the brightest emission-line amongst the observed, while the H$_2$ is the most compact emission, reaching $\approx$ 1$\farcs$0 preferably to the PA $\approx$ 50$^{\circ}$. 

The {\sc IFSCUBE} package \citep{rd2021} also provides the centroid velocity and velocity dispersion, that can be used to map the gas kinematics. In the second column of Fig.\,\ref{fig-flux-n} it is shown the centroid velocity fields after subtraction of the heliocentric systemic velocity of 3277 km s$^{-1}$. All velocity fields show blueshifts to the north-east and redshifts to the south-west, consistent with a rotation pattern, in agreement with the stellar velocity field. For the [Fe\,{\sc ii}]$\lambda$1.2570$\mu$m and Pa$\beta$ velocity fields we can see values ranging from -150 km s$^{-1}$ to 100 km s$^{-1}$, with smaller values for the H$_2$ velocity field, ranging from -100 km s$^{-1}$ to 75 km s$^{-1}$.

In the third column of Fig.\,\ref{fig-flux-n} we present the velocity dispersion maps. The Pa$\beta$ and H$_2$$\lambda$2.1218$\mu$m $\sigma$ maps are very similar, with values reaching $\approx$ 140 km s$^{-1}$ in regions to north-west and south-east for both. The [Fe\,{\sc ii}]$\lambda$1.2570$\mu$m $\sigma$ map, however, shows a clear increase in its values (reaching up to 200 km s$^{-1}$) in a bipolar structure in the north-west to south-east orientation. This increase in velocity dispersion is consistent with the interaction of the outflow -- primarily traced by the broad component -- with the interstellar medium, disturbing the gas in the disk and consequently raising the $\sigma$ values in intersection region.

%The Pa$\beta$ emission line flux extends up to 1.5$\arcsec$  in all directions, but been extended preferably from the north-east to the south-west. It is the brightest emission line amongst the observed ones and its 

%The H$_2$$\lambda$2.1218$\mu$m emission line flux extends up to 1.0$\arcsec$ preferably to the north-east to south-west direction. 

\subsection{The outflow component}

\begin{figure*}
	\centering
	\includegraphics[width=1.0\linewidth]{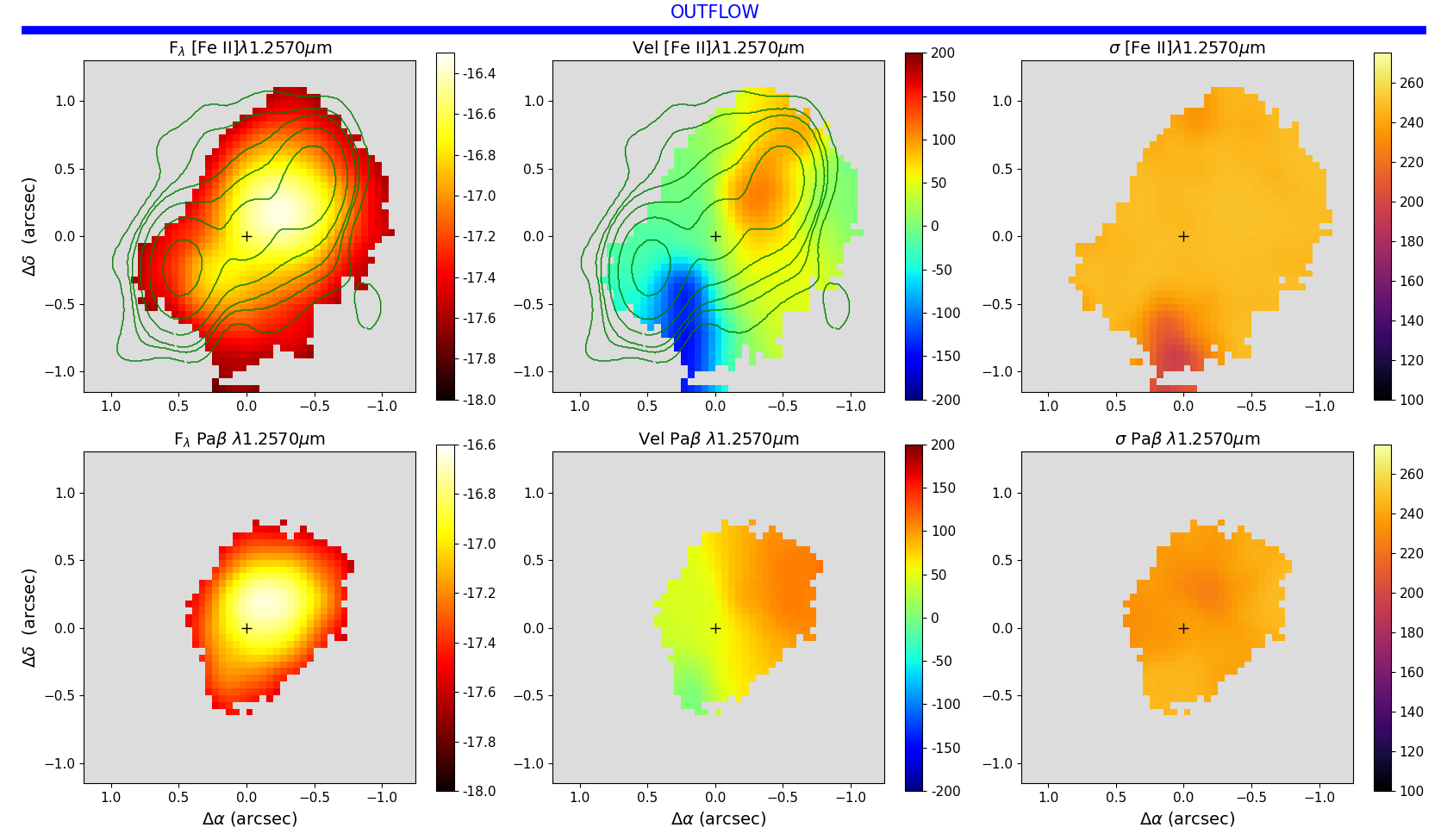}
	\caption{First column: flux distribution for the [Fe\,{\sc ii}]$\lambda$1.2570$\mu$m, Pa$\beta$ and H$_2$$\lambda$2.1218$\mu$m outflow (broad) component, respectively. The color bar shows the fluxes in logarithmic scale in units of erg s$^{-1}$ spaxel$^{-1}$. Second column: velocity maps for the same emission-lines of the first column. The color bar shows the velocity in units of km s$^{-1}$. Third column: $\sigma$ maps for the same emission-lines of the first column. The color bar shows the $\sigma$ values in units of km s$^{-1}$. The green contours on panels one and two are 8.4-Ghz radio observations by \citet{thean2000}. The central crosses marks the location of the peak of the continuum emission, the grey regions represent masked locations, where the amplitude of the corresponding line profile is smaller than 3 times the noise of the adjacent continuum.}
	\label{fig-flux-b}
\end{figure*}

In the first column of Fig.\,\ref{fig-flux-b} we present the flux distributions for the [Fe\,{\sc ii}]$\lambda$1.2570$\mu$m and Pa$\beta$; the H$_2$$\lambda$2.1218$\mu$m broad component is very faint and compact, therefore we do not show it. The [Fe\,{\sc ii}]$\lambda$1.2570$\mu$m emission-line flux distribution shows a very distinct distribution when compared with the narrow component, being approximately perpendicular to it, reaching 1$\farcs$0 ($\approx$ 230\,pc) to the north-west (region that coincides with the peak of the emission for the broad component) and 0$\farcs$8 ($\approx$ 180\,pc) to the south-east of the nucleus. The green contours represent the 8.4 GHz radio image from \citet{thean2000}, showing that the strongest radio emission is elongated in the southeast-northwest direction and is co-spatial with the broad component emission, clearly seen in the [Fe,{\sc ii}] emission. The Pa$\beta$ broad component, also distributed approximately perpendicular to the distribution of the narrow component, is more compact, reaching $\approx$ 0$\farcs$8 ($\approx$ 180\,pc) to the north-west, direction where it is located its emission peak.  

As in the case of the narrow component, centroid velocities fields were subtracted the heliocentric systemic velocity of 3277 km s$^{-1}$, and are shown in the second column of Fig.\,\ref{fig-flux-b}. The scenario is quite different from that rotation pattern seen in the narrow component, as we observe the [Fe\,{\sc ii}]$\lambda$1.2570$\mu$m velocity field showing a redshift spot to the north-west of the nucleus, with values close to 100 km s$^{-1}$ and a blueshifted spot to the south-east of the nucleus with values close to -150 km s$^{-1}$. The Pa$\beta$ velocity field shows only redshifts to the north-west region of the nucleus.

In the third column of Fig.\,\ref{fig-flux-b} the $\sigma$ maps show very disturbed gases with values close to 250 km s$^{-1}$ in all directions for the [Fe\,{\sc ii}] and Pa$\beta$ emission-lines.     

\subsection{Emission-line ratios}

\begin{figure}
	\centering
	\includegraphics[width=1.0\linewidth]{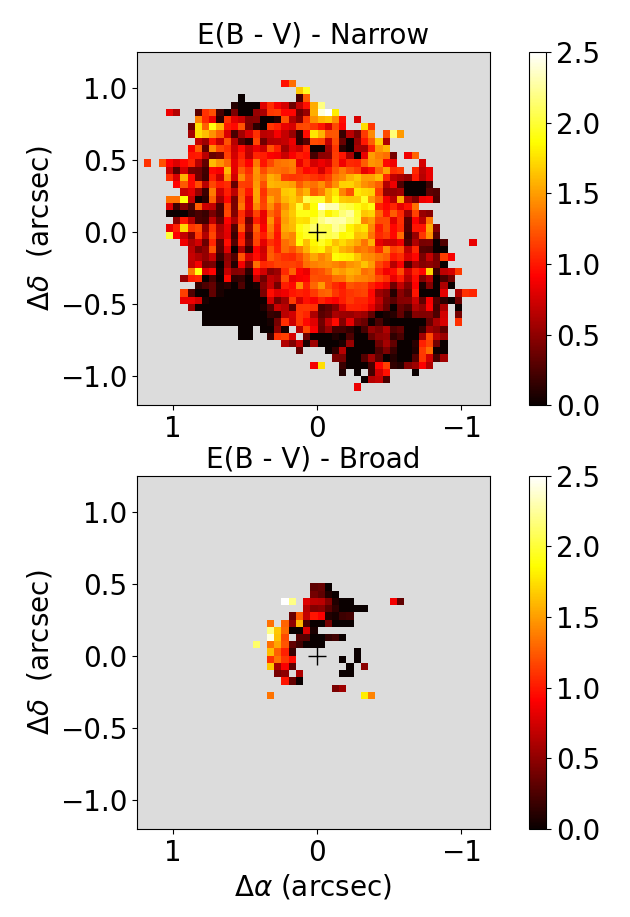}
	\caption{Top panel: Extinction E(B - V) obtained from the ratios of Pa$\beta$/Br$\gamma$ narrow components. Bottom panel: Extinction E(B - V) obtained from the ratios of Pa$\beta$/Br$\gamma$ broad components.}
	\label{ebv}
\end{figure}

In Fig.\,\ref{ebv} we show the color excess maps obtained using the equation:
\begin{equation}
    E(B-V) = 4.74 \log \left( \frac{5.88}{F_{\mathrm{Pa\beta}}/F_{\mathrm{Br\gamma}}}\right),
\end{equation}
where F$_{Pa\beta}$ and F$_{Br\gamma}$ are the fluxes of Pa$\beta$ and Br$\gamma$ emission-lines, respectively. We adopt the theoretical ratio between Pa$\beta$ and Br$\gamma$ of 5.88, corresponding to case B at the low-density limit \citep{osterbrock2006} for and electron temperature of T$_e =$ 10$^4$K and we use the reddening law of \cite{cardelli1989}. The map for the narrow component shows values reaching $\approx$2 mag in a region slightly north-west of the nucleus, with these values decreasing as we go further out in the FOV, with a median value of 0.86 for the whole E(B-V) map. For the broad component a more compact map is seen with an increase in the values to the north-west position, with the median value for the E(B-V) - broad of 0.36.

\begin{figure*}
	\centering
 \includegraphics[width=1.0\linewidth]{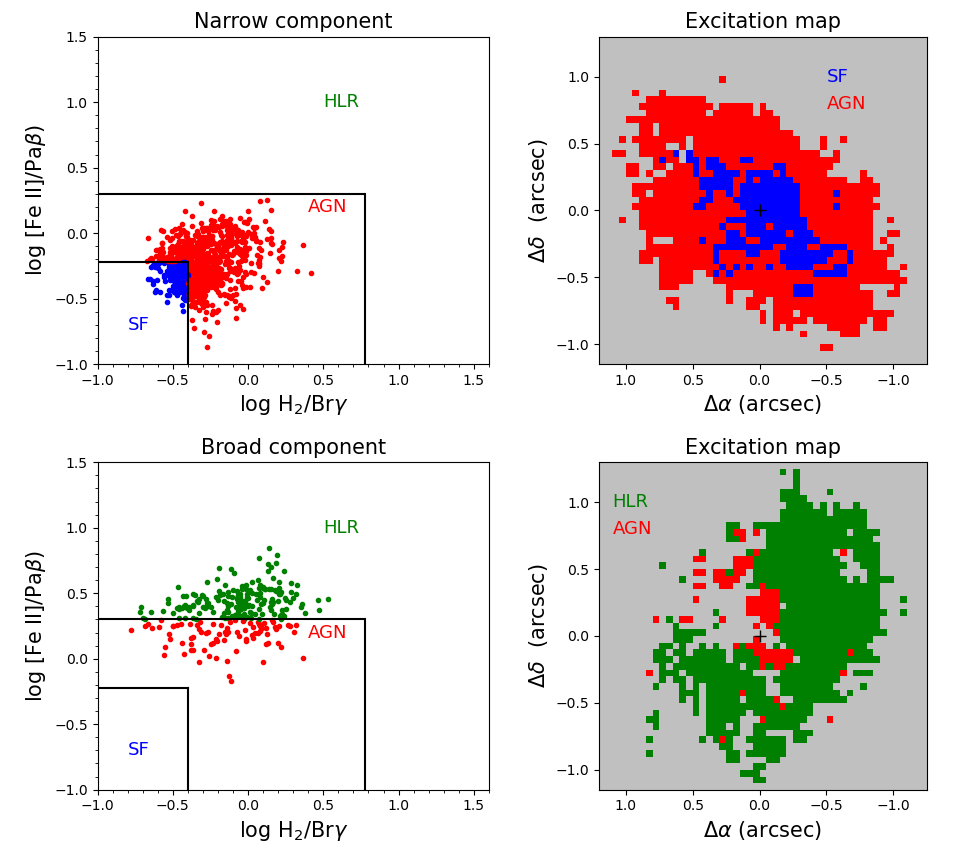}
	\caption{First row: [Fe\,{\sc ii}]$\lambda$1.2570$\mu$m/Pa$\beta$ vs. H$_2$$\lambda$2.1218$\mu$m/Br$\gamma$ diagnostic diagram (left) and corresponding excitation map (right) for the narrow component. Second row: the same for the broad component. The lines delineating the SF, AGN, and high line ratio (HLR) regions are from \citet{riffel2013a}.} 
	\label{diag-diag}
\end{figure*}

In Fig.\,\ref{diag-diag} we present the [Fe\,{\sc ii}]$\lambda$1.2570$\mu$m/Pa$\beta$ vs.H$_2$$\lambda$2.1218$\mu$m/Br$\gamma$ emission-line ratio diagnostic diagram, that can be used to investigate the origin of the [Fe\,{\sc ii}] and H$_2$ emission \citep{reunanen2002,ardila2004,ardila2005,riffel2013a,colina2015,riffel2020,riffel21}. The molecular hydrogen can be excited by distinct processes such as (i) fluorescence through absorption of soft-ultraviolet photons (912-1108 \AA) in the Lyman and Werner bands (these photons are present in star-forming (SF) regions and surrounding AGN \citep{black1987}), (ii) shocks \citep{hollenbach1989}, e.g. the interaction of a radio jet with the interstellar medium or from supernovae explosions \citep{larkin1998}, (iii) X-ray illumination by central AGN \citep{draine1990,maloney1996} or even (iv) UV radiation in dense clouds (densities ranging from 10$^{4}$ to 10$^{5}$ cm$^{-3}$ \citep{sternberg1989,davies2003}). Usually the UV fluorescence is regarded as a non-thermal process, while shocks and X-ray/UV heating are referred to as thermal processes, as these two produce distinct relative intensities between emission-lines of H$_2$, which can be used to determinate the dominant excitation mechanism.

The [Fe\,{\sc ii}] emission is due to shocks or X-ray excitation since the [Fe\,{\sc ii}] near-IR emission-lines are produced in the partially ionized gas phase and [Fe\,{\sc ii}]$\lambda$1.2570$\mu$m/Pa$\beta$ ratio is controlled by the ratio between the volumes of the partially to fully ionized gas \citep{mouri1990,mouri1993,ardila2005,riffel2008,storchi2009}. The partially ionized zones in the central region of galaxies originate from X-ray emission \citep{simpson1996} in AGN or shocks due to interaction of radio jets and gas outflows with ambient clouds \citep{forbes1993}.

%The [Fe\,{\sc ii}]$\lambda$1.2570$\mu$m/Pa$\beta$ versus H$_2$$\lambda$2.1218$\mu$m/Br$\gamma$ is the most commonly used diagram for this purpose. 
In the plots of Fig.\,\ref{diag-diag} to separate empirical limits for SF galaxies, AGN and high line ratio (HLR) objects, we used continuous vertical and horizontal lines. For the (1) SF galaxies the limits are [Fe\,{\sc ii}]$\lambda$1.2570$\mu$m/Pa$\beta$ $\le$ 0.6 and H$_2$$\lambda$2.1218$\mu$m/Br$\gamma$ $\le$ 0.4, for (2) AGN 0.6 $\le$ [Fe\,{\sc ii}]$\lambda$1.2570$\mu$m/Pa$\beta$ $\le$ 2 and 0.4 $\le$ H$_2$$\lambda$2.1218$\mu$m/Br$\gamma$ $\le$ 6 and for the (3) HLR (occupied by LINERs and Supernovae Remnants and shock dominated objects) the limits are [Fe\,{\sc ii}]$\lambda$1.2570$\mu$m/Pa$\beta$ $\ge$ 2 and H$_2$$\lambda$2.1218$\mu$m/Br$\gamma$ $\ge$ 6 \citep{rogerio2013}.

%The H$_2$ flux maps show also emission from locations where no ionized gas emission is seen and the H$_2$ emission is more likely been produced by shocks due to the interaction of a wide opening angle wind with the dense gas \citep[e.g.]{zaka2014,riffel2020} or by X-rays from the central AGN escaping through the dusty torus. The former is in agreement with shock models used to describe the H$_2$ molecule formation from luminous galaxies \citep{guillard2009}. For the latter one would expect the regions with H$_2$ emission and no ionized gas emission being located mainly perpendicular to the AGN ionization axis, which is not supported by the flux distributions described, for example, in \cite{riffel2021}, similarly in their study, there is [Fe {\sc ii}] emission in regions outside the AGN ionization cones, likely tracing the photodissociation regions at the cone edges.  

For the maps in Fig.\,\ref{diag-diag} the grey regions are masked locations, where the amplitude of the corresponding line profile is smaller than 3$\sigma$ of the adjacent continuum. For the narrow component (disk) we can see SF values close to the nucleus and AGN values on the outside with a preferential orientation from north-east to south-west. The broad component (outflow) shows AGN values in the nucleus and HLR values outside, in an orientation preferably from the south-east to the north-west.

\section{Discussion}

\subsection{Rotation velocity models}

The stellar velocity field (left-hand panel of Fig.\,\ref{stel-kin}) shows a rotation component, with the north-eastern side of the disc approaching and the south-western side receding with a velocity amplitude of about 100 km\,s$^{-1}$. In order to obtain parameters such as systemic velocity, orientation of the line of the nodes and inclination of the disc we fitted a rotation model for the stellar velocity field, assuming that it is rotating in a central potential \citep{bertola1991}. It is assumed, for this kinematic model \citep{vdk1978,bertola1991}, that the gas has circular orbits in a plane with the velocity field given by:
\begin{multline}
V_{\rm mod}(R,\Psi) = V_{\rm sys}+ \\
\,+ \frac{A\,R\,\cos(\Psi - \Psi_{0})\,\sin\,\theta \,\cos^{p}\,\theta}{(R^2)\,\left(\sin^2(\Psi - \Psi_0) + \cos^2\theta\,\cos^2(\Psi - \Psi_0) + c_{0}^{2}\,\cos^2\theta\right)^{p/2}}.
\end{multline}
where v$_{\rm sys}$ is the systemic velocity, A is the centroid velocity amplitude, $\Psi_0$ is the major axis position angle, $c_0$ is a concentration parameter, $\theta$ is the angle between the disc plane and the sky plane, $p$ is a model fitting parameter (where $p$ $\approx$ 1 for infinite masses in Plummer potential) and R and $\Psi$ are the coordinates of each spaxel in the plane of the sky. For these fits, we allowed all the parameters to vary, except A and $p$ that were bound to vary from 25 to 300 and 1 to 1.5 (limit values for galaxies \citep{bertola1991}) respectively.

\begin{figure*}
	\centering
	\includegraphics[width=1.0\linewidth]{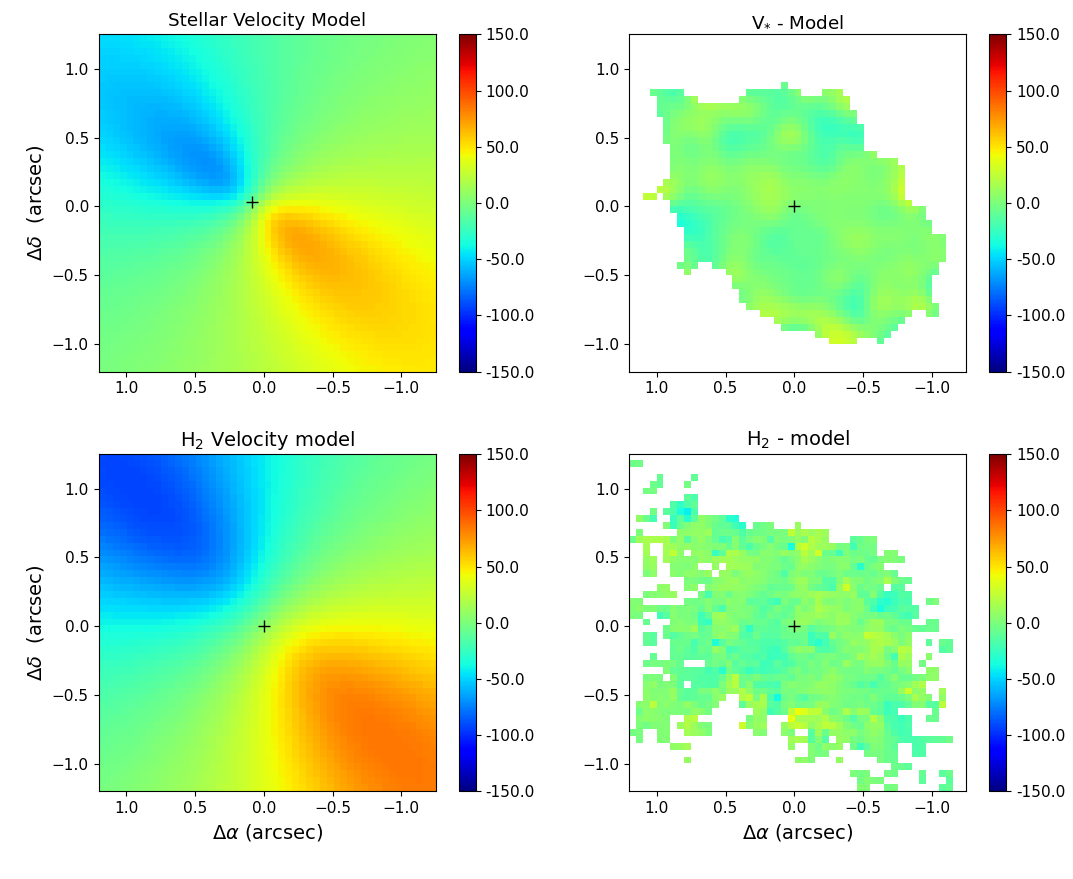}
	\caption{Rotating disc model fitted to the stars in the top left panel and its residuals in the right. The same for the H$_2 \lambda$2.1218$\mu$m velocity field in the bottom panels.}
	\label{fig-rot-stel-mol}
\end{figure*}

The best rotating disc model for the stars is shown in the top-left panel of Fig.\,\ref{fig-rot-stel-mol}. The residual map is in the top-right panel and does not show high values, indicating that the model reproduces well the observed velocity field. The corresponding kinematic parameters are shown in Table\,\ref{tabpar}. 

In Fig.\,\ref{fig-flux-n} we observe that all the velocity fields of the narrow components show a rotation pattern, but we chose the H$_2\lambda$2.1218$\mu$m and the Pa$\beta$ to fit the rotation model since they present lower velocity dispersion values, supporting that these species are more confined to the galaxy plane. In the bottom panels of Fig.\,\ref{fig-rot-stel-mol} we show the H$_2$ velocity model in the left and its residuals in the right panel.   

\begin{figure*}
	\centering
	\includegraphics[width=1.0\linewidth]{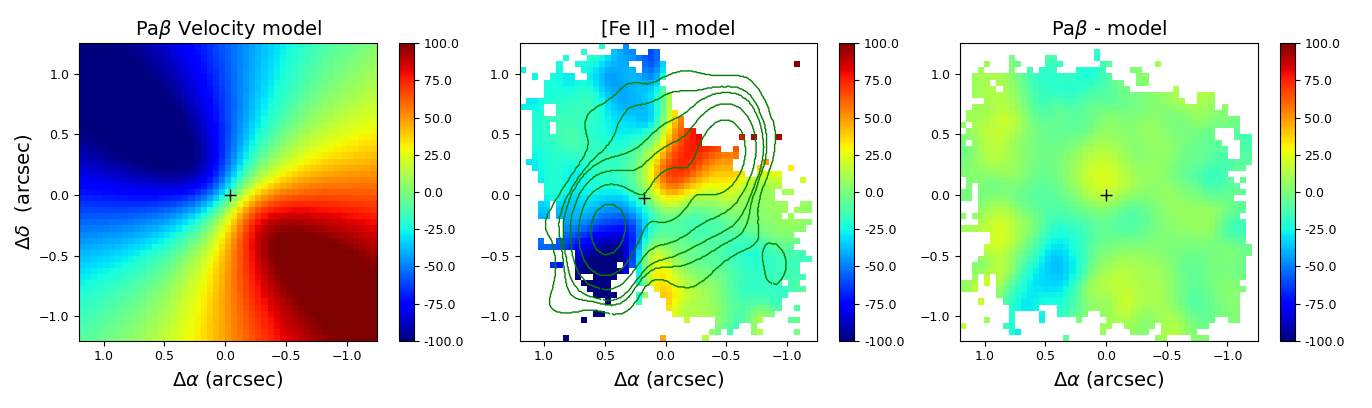}
	\caption{Rotating disc model fitted to the Pa$\beta$ velocity field, together with the residuals of its subtraction from the observed velocity fields of the narrow components of [Fe {\sc ii}]$\lambda$1.2570$\mu$m and Pa$\beta$. The green contours on panel two are 8.4-Ghz radio observations by \citet{thean2000}.}
	\label{fig-rot-model-ionized}
\end{figure*}

In Fig.\,\ref{fig-rot-model-ionized} we show the resulting Pa$\beta$ velocity model in the left panel and in the middle and right panel we show the [Fe\,{\sc ii}] and Pa$\beta$ narrow components residual maps (narrow component centroid velocity field of the emitting gas - model) respectively. The resulting parameters are shown in Table\,\ref{tabpar}.

\begin{table}
    \centering
    \begin{tabular}{l|c|c|c}
    \hline
       Parameter              & Stars &  H$_2$ & Pa$\beta$  \\
       \hline
       c$_0$ (arcsec)         & 6.1$\pm$1.1 &  14.7$\pm$2.2 & 8.7$\pm$1.3 \\
       $p$                    & 1.45$\pm$0.3  & 1.3$\pm$0.2 & 1.3$\pm$0.2 \\
       $\Psi_0$               & 51.1$\pm$5.2 &  44.6$^{\circ} \pm$4.0$^{\circ}$ & 49.9$^{\circ} \pm$4.5$^{\circ}$  \\
       $\theta$               & 55.5$\pm$8.3  & 56.5$^{\circ} \pm$9.0$^{\circ}$ & 46.4$^{\circ} \pm$9.4$^{\circ}$ \\
       v$_{sys}$ (km s$^{-1}$)& 3268$\pm$70 & 3258$\pm$65 & 3249$\pm$64 \\
       \hline
    \end{tabular}
    \caption{Values of the parameters c$_0$, $p$, $\Psi_0$, $\theta$ and v$_{sys}$ for the fit of the velocity model of the centroid velocity maps for the stars, H$_2 \lambda$2.1218$\mu$m and Pa$\beta$.}
    \label{tabpar}
\end{table}

We can see that the $c_0$ is larger for the resulting fit of the H$_2$ velocity field, which supports the idea that the H$_2$ emission is more compact and in the galaxy plane \citep{sb2010,riffel2011,riffel2013,diniz2015,riffel2020,schonell2019}. The $\Psi_{0}$ parameter is similar in all models, (51.1$^{\circ}$, 49.9$^{\circ}$ and 44.6$^{\circ}$ for the stars, Pa$\beta$ and H$_2$ respectively) and in good agreement with the orientation of the large scale disk of $sim$50$^\circ$ \citep{jarret03}. The systemic velocity found in the models range from 3249 km s$^{-1}$ for the Pa$\beta$ to 3268 km s$^{-1}$ for the stars, values that are in range from that found by \citep{theureau98} of 3277\,$\pm$\,5\,km\,s$^{-1}$. Although the residuals of the stars and H$_2$ are smooth and support a rotation pattern only, in the residual velocity map of the narrow component of [Fe\,{\sc ii}], we can clearly observe a red shift excess to the northwest and a blue shift excess to the southeast, supporting the idea that the outflow is interacting with the disk, as discussed in Sec.,\ref{results}. These red shift and blue shift excesses align with the 8.4\,GHz radio structure \citep{thean2000}, reinforcing the role of low and moderate-power jets as a feedback mechanism in local AGN \citep[e.g.][]{Nandi23,Girdhar24}.

\subsection{Mass-outflow rate and power}

\begin{figure}
	\centering
	\includegraphics[width=1.0\linewidth]{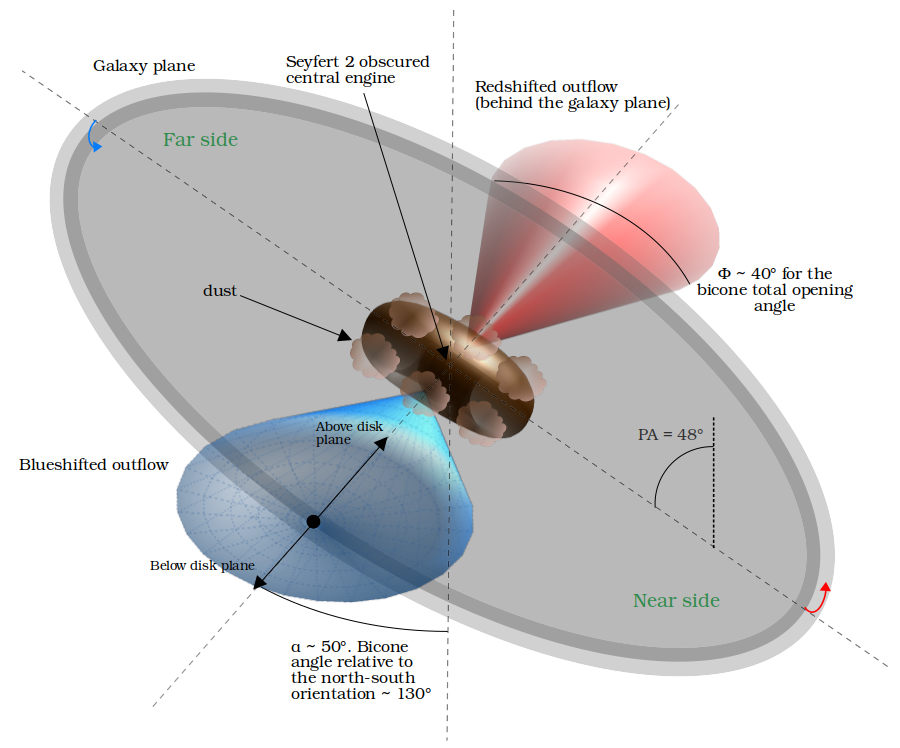}
	\caption{A scenario for the [Fe\,{\sc ii}]$\lambda$1.2570$\mu$m outflow, in which we can see blueshifts to the south-east (near-side) and redshifts to the north-west (far-side of the galaxy). The bicone outflow has an $\approx$ 20$^{\circ}$ aperture, making an angle of $\approx$ 130$^{\circ}$ with the north-south orientation. The nucleus is obscured by the dusty torus, hence the Seyfert 2 classification, as we can see by the nuclear spectra.}
	\label{fig-outflow}
\end{figure}

We identified two kinematic components in the gas: one dominated by the disk with a contribution of the outflow, observed as a narrow component in the emission lines, and another associated with the outflow, seen as a broad component in the emission lines. In the $\sigma$ map of the [Fe\,{\sc ii}]$\lambda$1.2570$\mu$m narrow component shown in Fig.\,\ref{fig-flux-n}, we observe a region of enhanced $\sigma$ values, extending from north-west to south-east, and suggesting an interaction between the outflow and the gas disk. This interaction is further supported by the residual velocity map shown in Fig.\,\ref{fig-rot-model-ionized}, which reveals excess redshifts in the north-west and blueshifts in the south-east within the same region (NW-SE), aligning with the orientation of the region of enhanced $\sigma$ values.  Finally, we observe that the extent of the broad component of the ionized emission closely follows the radio structure (Fig.\,\ref{fig-flux-b}), agreeing with recent studies \citep[e.g.][]{nandi2023,girdhar2024} in which low-power jets provide an important source of feedback in nearby galaxies. With all this in mind and using the previously obtained disk orientation parameters, we can constrain the geometry of the ionized gas outflows in the inner 300$\times$300 pc$^2$ of NGC\:1125. The close correlation between the line emission, kinematics, and the radio structure suggests that the outflow is driven by the radio jet, which is oriented along PA $\approx$ 130$^{\circ}$ \citep{thean2000}. We assume an opening angle for the bipolar outflow of 40$^{\circ}$, as estimated directly from the velocity field of the broad component for the ionized gas. The disk major axis is orientated along $\Psi_0\approx$ 48$^{\circ}$ and it is inclined relative to the plane of the sky by $\theta\approx$ 50$^{\circ}$, as obtained by the modeling of the narrow-component velocity fields. We cannot infer the exact angle of the bicone relative to the line of sight, but we can estimate a minimum value of 20$^{\circ}$, since this is a Seyfert 2 AGN, and the central engine cannot be observed directly. With a minimum limit of 20$^{\circ}$, we would be observing the front wall of the cone. The maximum value is 40$^{\circ}$, as this is the angle between the galaxy disk and the line of sight, providing the limit for observing the outflow along the disk plane, as it interacts with the gas in the disk. The geometry of the outflow is reproduced in an schematic way in Fig.\,\ref{fig-outflow}.

The estimate for the mass outflow rate in the ionized gas depends on the geometry of the structure. Therefore, \cite{lutz2020} have shown that for a bipolar structure, such as in NGC\,1125, one can adopt: 
\begin{equation}
    \dot{M}_{\mathrm{out}}=3\,\frac{M_{\mathrm{out}}\,v_{\mathrm{out}}}{R_{\mathrm{out}}},
    \label{outflow}
\end{equation}
where $M_{\mathrm{out}}$ is calculated from \citep{osterbrock2006,sb2009}:

\begin{equation}
    \left(\frac{M_{\mathrm{H\,II}}}{\mathrm{M_{\sun}}}\right)= 5.1\,\times\,10^{18} \left( \frac{F_{\mathrm{Pa\beta}}}{\mathrm{erg\, cm^{-2}\,s^{-1}}} \right) \left(\frac{D}{\mathrm{Mpc}} \right)^2\left(\frac{N_\mathrm{e}}{\mathrm{cm^{-3}}} \right)^{-1},
    \label{emout}
\end{equation}
and $F_{\mathrm{Pa\beta}}$ is the total flux for the broad component of Pa$\mathrm{\beta}$, $D$ is the distance to the galaxy and $N_e$ is the electron density of the outflow. The total flux for the broad component of Pa$\beta$ after extinction correction is $F_{\mathrm{Pa\beta}}$ = 5.2$\times$10$^{-15}$ erg\,s$^{-1}$\,cm$^{-2}$, using  $A_V=1.12$ mag, the extinction law of \citet{cardelli1989} and adopting an electron density of $N_e=$500 cm$^{-3}$, which is a typical value for AGN hosts \citep[e.g.][]{dors14,dors20,Freitas18,kakkad18}. The method used to estimate the density of ionized outflows is one of the main sources of uncertainty to estimate mass outflow rates \citep{baron2019,davies2020,revalski2022} as a wide range of densities (from 10$^2$ to 10$^4$ cm$^{-3}$) have been adopted in the last decade \citep[e.g.][]{liu2013,diniz2019,kakkad2020} when it cannot be directly estimated from the data used. Alternative diagnostics \citep{rose2018,santoro2020,holden2023a,holden2023b} predict densities that are one order of magnitude higher than $N_e=$500 cm$^{-3}$ and since the mass of ionized gas is inversely proportional to the electron density (see Eq. 4), the use of [S\:{\sc ii}]-based $N_e$ may lead to overestimated mass outflow rates \citep{davies2020}.

We observe the portion of the outflow above the disk plane, with blueshifts to the south (near side of the galaxy) and redshifts to the north (far side of the galaxy). The values for the projected velocities are calculated from  $v_{\rm out} = \frac{v}{\sin \gamma}$ km s$^{-1}$, where $\gamma$ is the orientation of the bicone relative to the line of sight and $v$ is defined as the average projected velocity over the region dominated by outflow, weighted by flux of the broad component, calculated by \citep{riffel2023}:

\begin{equation}
\label{eq-vel}
    v =  \frac{\langle |V|\,F_{\rm [Fe\,II]} \rangle}{\langle F_{\rm [Fe\,II]} \rangle},
\end{equation}
where $V$ are the velocities of the [Fe\,{\sc ii}] broad component in each spaxel and $F_{[Fe\,II]}$ are their fluxes. From Eq.\,\ref{eq-vel} we obtain that the projected velocity of the outflowing gas is 125 km s$^{-1}$, therefore the values for the velocity of the outflow are $v_{\rm out}\approx$ 195 and 370 km s$^{-1}$, using the minimum and maximum $\gamma$ values of 20$^{\circ}$ and 40$^{\circ}$, as discussed above.

The radius of the bulk of the outflow $R_{\rm out}$ is defined as \citep{riffel2023}:
\begin{equation}
\label{eq-r}
    R_{\mathrm{out}} =  \frac{\langle R\,F_{\mathrm{Pa\beta}} \rangle}{\langle F_{\mathrm{Pa\beta}} \rangle},
\end{equation}
where $R$ are the distances of  each spaxel from the galaxy's nucleus, with broad line detected. The value for $R_{\rm out}$ then calculated as 115\,pc (0.5$\arcsec$).

Using the values discussed above in the equation \ref{outflow}, with values of $\gamma$ been in a maximum range of 40$^{\circ}$ and a minimum range of 20$^{\circ}$, we find upper and lower values for the mass outflow rate of 1.1\,$\pm$\,0.2\,M$_{\sun}$ yr$^{-1}$ and 0.6\,$\pm$\,0.1\,M$_{\sun}$ yr$^{-1}$, respectively. These values are within the range commonly reported for AGNs of similar luminosities. These values are 3-5$\times$ higher than those found in \citet{riffel2023}, of $\log \dot{M}_{\mathrm{out}}/[{\rm M_\odot yr^{-1}}] = -0.64\pm0.24$, for NGC\:1125. However, as discussed by these authors, their estimates only include outflow-dominated spaxels, since their method relies on using the line width to estimate the outflow velocity, without performing the deblending of the outflow and disk components. 
%These values are in good agreement with previous studies of outflows in the ionized gas as in \cite{schonell2019} of 6.8 M$_{\sun}$ yr$^{-1}$ for NGC\,5548. In \citep{bianchin2021}, the outflows in ionized gas for 4 AGN range from 5$\times$10$^{-3}$ to 12.5 M$_{\sun}$ yr$^{-1}$, while in \cite{riffel2023} the ionized outflow values of 33 AGN were calculated and are in the range of 1.3$\times$10$^{-3}$ to 20 M$_{\sun}$ yr$^{-1}$. In the same study, NGC\,1125 ionized outflow was calculated to be 0.22 M$_{\sun}$ yr$^{-1}$. }

From the outflow rate, one can derive the outflow kinematic power as \citep{sb2010,schonell2014,schonell2019}:
 \begin{equation}
     \dot{E}_{\mathrm{kin}}\approx \frac{\dot{M}_{\mathrm{out}}}{2}(v^{2}_{\mathrm{out}}+3\sigma^{2}_{out}),
 \end{equation}
where $v_{\rm out}$ is the velocity of the outflowing gas and $\sigma_{\rm out}$ is its velocity dispersion. Using $v_{\rm out}$ = 195 and 370 km s$^{-1}$ (as discussed above) and $\sigma$ = 240 km s$^{-1}$ (median value of the $\sigma$ map from the broad component), we obtained a maximum value of $\dot{E}\approx$ 1.1 $\times$ 10$^{41}$ erg s$^{-1}$ and a minimum value of 3.9 $\times$ 10$^{40}$ erg s$^{-1}$. %These values are in good agreement with those found for AGN \citep{morganti2005}, being close to those found by \cite{riffel2011b,schonell2014,schonell2019} of $\dot{E}\approx$ 5.7 $\times$ 10$^{41}$ for Mrk\,1157, $\dot{E}\approx$ 2.9 $\times$ 10$^{41}$ for Mrk\,766 and $\dot{E}\approx$ 2.1 $\times$ 10$^{41}$ for NGC\,5548, respectively. These values are also close to those found by \cite{ms2011} that range from 0.6 to 50 $\times$ 10$^{41}$ erg s$^{-1}$.}
In order to compare the kinematic power of the outflow with the bolometric luminosity of the AGN of NGC\,1125, which can be obtained from the hard X-ray (14-195 keV) luminosity, $L_{\rm X}$, by log\,$L_{\rm bol}$\,=\,0.0378(log $L_{\rm X}$)$^{2}$\,-\,2.03 log\,L$_X$\,+\,61.6 \citep{ichikawa2017}, we used log\,$L_{\rm X}$\,=\,42.64 \citep{riffel2021}, obtaining L$_{bol}$\,=\,5.9 $\times$ 10$^{43}$ erg s$^{-1}$. Therefore, the kinematic power of the outflow represents 0.07\% and 0.2\% of the bolometric luminosity of the AGN. In order to suppress star formation, the models require a minimum coupling efficiency ($\epsilon_{f}$) for the AGN feedback in the range of 0.5 to 20 per cent \citep{dimatteo2005,hopkins2010,dubois2014,schaye2015,weinberger2017}. However, it is unlikely that all the outflow energy becomes kinetic power, therefore, a direct comparison between observed $\dot{E}_{\rm kin}$/$L_{\rm bol}$ and predicted $\epsilon_{f}$ is not  straightforward. From numeric simulations \citet{richings2018} indicated that kinetic energy of the outflows represents $<$ 20 per cent of the total emitted outflow energy. If we look in the hot molecular gas outflows, we find that they are 2 orders of magnitude lower than those of the ionized gas, ergo not being powerful enough to suppress star formation. On the other hand, AGN outflows are seen in multiple gas phases and the kinetic power of dense cold molecular outflows are expected to be larger. Therefore, even that ionized outflows are not enough to suppress star formation, we can not rule out the possibility that they are important mechanisms in the evolution of the galaxies.

\section{Conclusions}
We mapped the gas distribution and kinematics, as well as the stellar kinematics, in the inner approximately 300 pc of the Seyfert 2 galaxy NGC\:1125 using integral field spectroscopy with Gemini NIFS in the J- and K-bands. The main findings of this study are as follows:

\begin{itemize}

\item The stellar velocity field is dominated by rotation and the velocity dispersion map presents a ring of low $\sigma$ with a radius of $\approx$ 100\,pc, attributed to intermediate-age stars that dominates the light in the central region of this galaxy;

\item  The emission-line flux distributions of molecular hydrogen H$_2$ and low-ionization gas are extended to at least $\approx$ 230\,pc from the nucleus;

\item There are two kinematic components in most emission lines: a narrow ($\sigma$ $\approx$ 80 km s$^{-1}$), that we attribute to rotation in the plane of the galaxy and a broad ($\sigma$ $\approx$ 240 km s$^{-1}$), attributed to an outflow; 

\item [Fe\,{\sc ii}]$\lambda$1.2570$\mu$m, Pa$\beta$ and H$_2$$\lambda$2.1218$\mu$m narrow component emission-line fluxes are most extended along PA$\approx$ 50$^{\circ}$, which is close to the PA of the line of the nodes of the gas kinematics (49.87$\pm$4.48 for the Pa$\beta$ rotation model);

\item The [Fe\,{\sc ii}]$\lambda$1.2570$\mu$m and Pa$\beta$ broad component are most extended along the perpendicular direction of the line of the nodes of the gas kinematics;

\item The narrow component excitation diagnostic diagrams show a mixture of AGN and SF values, where the AGN values extend up to 1$\arcsec$ in the north-east to south-west direction, and the SF values are closer to the nucleus;

\item The broad component excitation diagnostic diagrams shows AGN values in the nucleus and HLR values in the north-west to south-east direction, close to the orientation of the radio structure as well as the corresponding increase in the $\sigma$ map for the [Fe\,{\sc ii}] narrow component, supporting and a contribution from shocks to the gas excitation;

\item The Pa$\beta$ and H$_2$ narrow component kinematics are dominated by rotation with the H$_2$ kinematics been more compact with an amplitude of $\approx$ 100 km s$^{-1}$ and low velocity dispersion ($\approx$ 80 km s$^{-1}$);

\item The [Fe\,{\sc ii}] broad component kinematic is also dominated by rotation, but the high velocity dispersion map and the residual map from the model, show in addition an outflow component in the north-west to south-east position angle, well described by Fig.\,\ref{outflow};

\item The mass outflow rates are estimated to be between 0.6 to 1.1 M$_{\sun}$ yr$^{-1}$ and the power of the outflow $\approx$ 0.4 to 1.1$\times$10$^{41}$ erg s$^{-1}$, which represents 0.07\% and 0.2\% of the bolometric luminosity of the AGN.

\item The clear connection of the shock ionized outflow with the relatively low-luminosity radio source add to the growing evidence that low-power jets provide an important source of feedback in nearby galaxies.

\end{itemize}

\section*{Acknowledgements}
We would like to thank an anonymous referee for their contributions, which greatly improved this manuscript. RAR acknowledges the support from Conselho Nacional de Desenvolvimento Cient\'ifico e Tecnol\'ogico (CNPq; Proj. 303450/2022-3, 403398/2023-1, \& 441722/2023-7), Funda\c c\~ao de Amparo \`a pesquisa do Estado do Rio Grande do Sul (FAPERGS; Proj. 21/2551-0002018-0), and Coordena\c c\~ao de Aperfei\c coamento de Pessoal de N\'ivel Superior (CAPES;  Proj. 88887.894973/2023-00). RR acknowledges support from CNPq (Proj. 311223/2020-6,  304927/2017-1, 400352/2016-8, and  404238/2021-1), FAPERGS (Proj. 19/1750-2 and 24/2551-0001282-6) and (CAPES, Proj. Proj. 88887.894973/2023-00). Based on observations obtained at the Gemini Observatory, which is operated by the Association of Universities for Research in Astronomy, Inc., under a cooperative agreement with the NSF on behalf of the Gemini partnership: the National Science Foundation (United States), National Research Council (Canada), CONICYT (Chile), Ministerio de Ciencia, Tecnolog\'{i}a e Innovaci\'{o}n Productiva (Argentina), Minist\'{e}rio da Ci\^{e}ncia, Tecnologia e Inova\c{c}\~{a}o (Brazil), and Korea Astronomy and Space Science Institute (Republic of Korea).  This research has made use of NASA's Astrophysics Data System Bibliographic Services. This research has made use of the NASA/IPAC Extragalactic Database (NED), which is operated by the Jet Propulsion Laboratory, California Institute of Technology, under contract with the National Aeronautics and Space Administration.

\section*{DATA AVAILABILITY}
The data used in this paper are available in the Gemini Science Archive at: https://archive.gemini.edu/searchform. The processed data used in this paper will be shared on reasonable request to the corresponding author.

\bibliographystyle{mnras}
\bibliography{references} % if your bibtex file is called example.bib

\begin{thebibliography}{}
\makeatletter
\relax
\def\mn@urlcharsother{\let\do\@makeother \do\$\do\&\do\#\do\^\do\_\do\%\do\~}
\def\mn@doi{\begingroup\mn@urlcharsother \@ifnextchar [ {\mn@doi@} {\mn@doi@[]}}
\def\mn@doi@[#1]#2{\def\@tempa{#1}\ifx\@tempa\@empty \href {http://dx.doi.org/#2} {doi:#2}\else \href {http://dx.doi.org/#2} {#1}\fi \endgroup}
\def\mn@eprint#1#2{\mn@eprint@#1:#2::\@nil}
\def\mn@eprint@arXiv#1{\href {http://arxiv.org/abs/#1} {{\tt arXiv:#1}}}
\def\mn@eprint@dblp#1{\href {http://dblp.uni-trier.de/rec/bibtex/#1.xml} {dblp:#1}}
\def\mn@eprint@#1:#2:#3:#4\@nil{\def\@tempa {#1}\def\@tempb {#2}\def\@tempc {#3}\ifx \@tempc \@empty \let \@tempc \@tempb \let \@tempb \@tempa \fi \ifx \@tempb \@empty \def\@tempb {arXiv}\fi \@ifundefined {mn@eprint@\@tempb}{\@tempb:\@tempc}{\expandafter \expandafter \csname mn@eprint@\@tempb\endcsname \expandafter{\@tempc}}}

\bibitem[\protect\citeauthoryear{{Barbosa}, {Storchi-Bergmann}, {McGregor}, {Vale}  \& {Rogemar Riffel}}{{Barbosa} et~al.}{2014}]{barbosa2014}
{Barbosa} F.~K.~B.,  {Storchi-Bergmann} T.,  {McGregor} P.,  {Vale} T.~B.,   {Rogemar Riffel} A.,  2014, \mn@doi [\mnras] {10.1093/mnras/stu1637}, \href {https://ui.adsabs.harvard.edu/abs/2014MNRAS.445.2353B} {445, 2353}

\bibitem[\protect\citeauthoryear{Baron \& Netzer}{Baron \& Netzer}{2019}]{baron2019}
Baron D.,  Netzer H.,  2019, \mn@doi [Monthly Notices of the Royal Astronomical Society] {10.1093/mnras/stz1070}, 486, 4290

\bibitem[\protect\citeauthoryear{{Bertola}, {Bettoni}, {Danziger}, {Sadler}, {Sparke}  \& {de Zeeuw}}{{Bertola} et~al.}{1991}]{bertola1991}
{Bertola} F.,  {Bettoni} D.,  {Danziger} J.,  {Sadler} E.,  {Sparke} L.,   {de Zeeuw} T.,  1991, \mn@doi [\apj] {10.1086/170058}, \href {https://ui.adsabs.harvard.edu/abs/1991ApJ...373..369B} {373, 369}

\bibitem[\protect\citeauthoryear{{Black} \& {van Dishoeck}}{{Black} \& {van Dishoeck}}{1987}]{black1987}
{Black} J.~H.,  {van Dishoeck} E.~F.,  1987, \mn@doi [\apj] {10.1086/165740}, \href {https://ui.adsabs.harvard.edu/abs/1987ApJ...322..412B} {322, 412}

\bibitem[\protect\citeauthoryear{{Cano-D{\'\i}az}, {Maiolino}, {Marconi}, {Netzer}, {Shemmer}  \& {Cresci}}{{Cano-D{\'\i}az} et~al.}{2012}]{cano2012}
{Cano-D{\'\i}az} M.,  {Maiolino} R.,  {Marconi} A.,  {Netzer} H.,  {Shemmer} O.,   {Cresci} G.,  2012, \mn@doi [\aap] {10.1051/0004-6361/201118358}, \href {https://ui.adsabs.harvard.edu/abs/2012A&A...537L...8C} {537, L8}

\bibitem[\protect\citeauthoryear{{Cappellari} \& {Emsellem}}{{Cappellari} \& {Emsellem}}{2004}]{cappellari2004}
{Cappellari} M.,  {Emsellem} E.,  2004, \mn@doi [\pasp] {10.1086/381875}, \href {https://ui.adsabs.harvard.edu/abs/2004PASP..116..138C} {116, 138}

\bibitem[\protect\citeauthoryear{{Cardelli}, {Clayton}  \& {Mathis}}{{Cardelli} et~al.}{1989}]{cardelli1989}
{Cardelli} J.~A.,  {Clayton} G.~C.,   {Mathis} J.~S.,  1989, \mn@doi [\apj] {10.1086/167900}, \href {https://ui.adsabs.harvard.edu/abs/1989ApJ...345..245C} {345, 245}

\bibitem[\protect\citeauthoryear{{Chambers} et~al.,}{{Chambers} et~al.}{2016}]{chambers2016}
{Chambers} K.~C.,  et~al., 2016, \mn@doi [arXiv e-prints] {10.48550/arXiv.1612.05560}, \href {https://ui.adsabs.harvard.edu/abs/2016arXiv161205560C} {p. arXiv:1612.05560}

\bibitem[\protect\citeauthoryear{{Cicone} et~al.,}{{Cicone} et~al.}{2014}]{cicone2012}
{Cicone} C.,  et~al., 2014, \mn@doi [\aap] {10.1051/0004-6361/201322464}, \href {https://ui.adsabs.harvard.edu/abs/2014A&A...562A..21C} {562, A21}

\bibitem[\protect\citeauthoryear{{Colina} et~al.,}{{Colina} et~al.}{2015}]{colina2015}
{Colina} L.,  et~al., 2015, \mn@doi [\aap] {10.1051/0004-6361/201425567}, \href {https://ui.adsabs.harvard.edu/abs/2015A&A...578A..48C} {578, A48}

\bibitem[\protect\citeauthoryear{Dahmer-Hahn et~al.,}{Dahmer-Hahn et~al.}{2019}]{dahmer2019}
Dahmer-Hahn L.~G.,  et~al., 2019, \mn@doi [Monthly Notices of the Royal Astronomical Society] {10.1093/mnras/stz2453}, 489, 5653

\bibitem[\protect\citeauthoryear{{Davies}, {Sternberg}, {Lehnert}  \& {Tacconi-Garman}}{{Davies} et~al.}{2003}]{davies2003}
{Davies} R.~I.,  {Sternberg} A.,  {Lehnert} M.,   {Tacconi-Garman} L.~E.,  2003, \mn@doi [\apj] {10.1086/378634}, \href {https://ui.adsabs.harvard.edu/abs/2003ApJ...597..907D} {597, 907}

\bibitem[\protect\citeauthoryear{Davies et~al.,}{Davies et~al.}{2020}]{davies2020}
Davies R.,  et~al., 2020, \mn@doi [Monthly Notices of the Royal Astronomical Society] {10.1093/mnras/staa2413}, 498, 4150

\bibitem[\protect\citeauthoryear{{Di Matteo}, {Springel}  \& {Hernquist}}{{Di Matteo} et~al.}{2005}]{dimatteo2005}
{Di Matteo} T.,  {Springel} V.,   {Hernquist} L.,  2005, \mn@doi [\nat] {10.1038/nature03335}, \href {https://ui.adsabs.harvard.edu/abs/2005Natur.433..604D} {433, 604}

\bibitem[\protect\citeauthoryear{{Diniz}, {Riffel}, {Storchi-Bergmann}  \& {Winge}}{{Diniz} et~al.}{2015}]{diniz2015}
{Diniz} M.~R.,  {Riffel} R.~A.,  {Storchi-Bergmann} T.,   {Winge} C.,  2015, \mn@doi [\mnras] {10.1093/mnras/stv1694}, \href {https://ui.adsabs.harvard.edu/abs/2015MNRAS.453.1727D} {453, 1727}

\bibitem[\protect\citeauthoryear{Diniz, Riffel, Storchi-Bergmann  \& Riffel}{Diniz et~al.}{2019}]{diniz2019}
Diniz M.~R.,  Riffel R.~A.,  Storchi-Bergmann T.,   Riffel R.,  2019, \mn@doi [Monthly Notices of the Royal Astronomical Society] {10.1093/mnras/stz1329}, 487, 3958

\bibitem[\protect\citeauthoryear{{Dors}, {Cardaci}, {H{\"a}gele}  \& {Krabbe}}{{Dors} et~al.}{2014}]{dors14}
{Dors} O.~L.,  {Cardaci} M.~V.,  {H{\"a}gele} G.~F.,   {Krabbe} {\^A}.~C.,  2014, \mn@doi [\mnras] {10.1093/mnras/stu1218}, \href {https://ui.adsabs.harvard.edu/abs/2014MNRAS.443.1291D} {443, 1291}

\bibitem[\protect\citeauthoryear{{Dors}, {Maiolino}, {Cardaci}, {H{\"a}gele}, {Krabbe}, {P{\'e}rez-Montero}  \& {Armah}}{{Dors} et~al.}{2020}]{dors20}
{Dors} O.~L.,  {Maiolino} R.,  {Cardaci} M.~V.,  {H{\"a}gele} G.~F.,  {Krabbe} A.~C.,  {P{\'e}rez-Montero} E.,   {Armah} M.,  2020, \mn@doi [\mnras] {10.1093/mnras/staa1781}, \href {https://ui.adsabs.harvard.edu/abs/2020MNRAS.tmp.1925D} {}

\bibitem[\protect\citeauthoryear{{Draine} \& {Woods}}{{Draine} \& {Woods}}{1990}]{draine1990}
{Draine} B.~T.,  {Woods} D.~T.,  1990, \mn@doi [\apj] {10.1086/169358}, \href {https://ui.adsabs.harvard.edu/abs/1990ApJ...363..464D} {363, 464}

\bibitem[\protect\citeauthoryear{{Dubois} et~al.,}{{Dubois} et~al.}{2014}]{dubois2014}
{Dubois} Y.,  et~al., 2014, \mn@doi [\mnras] {10.1093/mnras/stu1227}, \href {https://ui.adsabs.harvard.edu/abs/2014MNRAS.444.1453D} {444, 1453}

\bibitem[\protect\citeauthoryear{{Fabian}}{{Fabian}}{2012}]{fabian2012}
{Fabian} A.~C.,  2012, \mn@doi [\araa] {10.1146/annurev-astro-081811-125521}, \href {https://ui.adsabs.harvard.edu/abs/2012ARA&A..50..455F} {50, 455}

\bibitem[\protect\citeauthoryear{{Ferrarese} \& {Ford}}{{Ferrarese} \& {Ford}}{2005}]{ferrarese2005}
{Ferrarese} L.,  {Ford} H.,  2005, \mn@doi [\ssr] {10.1007/s11214-005-3947-6}, \href {https://ui.adsabs.harvard.edu/abs/2005SSRv..116..523F} {116, 523}

\bibitem[\protect\citeauthoryear{{Flewelling}}{{Flewelling}}{2016}]{flewelling2016}
{Flewelling} H.,  2016, in American Astronomical Society Meeting Abstracts \#227. p. 144.25

\bibitem[\protect\citeauthoryear{{Forbes} \& {Ward}}{{Forbes} \& {Ward}}{1993}]{forbes1993}
{Forbes} D.~A.,  {Ward} M.~J.,  1993, \mn@doi [\apj] {10.1086/173221}, \href {https://ui.adsabs.harvard.edu/abs/1993ApJ...416..150F} {416, 150}

\bibitem[\protect\citeauthoryear{{Freitas} et~al.,}{{Freitas} et~al.}{2018}]{Freitas18}
{Freitas} I.~C.,  et~al., 2018, \mn@doi [\mnras] {10.1093/mnras/sty303}, \href {https://ui.adsabs.harvard.edu/abs/2018MNRAS.476.2760F} {476, 2760}

\bibitem[\protect\citeauthoryear{Gallagher, Maiolino, Belfiore, Drory, Riffel  \& Riffel}{Gallagher et~al.}{2019}]{gallagher2019}
Gallagher R.,  Maiolino R.,  Belfiore F.,  Drory N.,  Riffel R.,   Riffel R.,  2019, \mn@doi [Monthly Notices of the Royal Astronomical Society] {10.1093/mnras/stz564}, 485, 3409

\bibitem[\protect\citeauthoryear{{Girdhar} et~al.,}{{Girdhar} et~al.}{2024a}]{Girdhar24}
{Girdhar} A.,  et~al., 2024a, \mn@doi [\mnras] {10.1093/mnras/stad3453}, \href {https://ui.adsabs.harvard.edu/abs/2024MNRAS.527.9322G} {527, 9322}

\bibitem[\protect\citeauthoryear{{Girdhar} et~al.,}{{Girdhar} et~al.}{2024b}]{girdhar2024}
{Girdhar} A.,  et~al., 2024b, \mn@doi [\mnras] {10.1093/mnras/stad3453}, \href {https://ui.adsabs.harvard.edu/abs/2024MNRAS.527.9322G} {527, 9322}

\bibitem[\protect\citeauthoryear{{Harrison}}{{Harrison}}{2017}]{harrison2017}
{Harrison} C.~M.,  2017, \mn@doi [Nature Astronomy] {10.1038/s41550-017-0165}, \href {https://ui.adsabs.harvard.edu/abs/2017NatAs...1E.165H} {1, 0165}

\bibitem[\protect\citeauthoryear{{Harrison} \& {Ramos Almeida}}{{Harrison} \& {Ramos Almeida}}{2024}]{harrison2024}
{Harrison} C.~M.,  {Ramos Almeida} C.,  2024, \mn@doi [Galaxies] {10.3390/galaxies12020017}, \href {https://ui.adsabs.harvard.edu/abs/2024Galax..12...17H} {12, 17}

\bibitem[\protect\citeauthoryear{{Holden} \& {Tadhunter}}{{Holden} \& {Tadhunter}}{2023}]{holden2023a}
{Holden} L.~R.,  {Tadhunter} C.~N.,  2023, \mn@doi [\mnras] {10.1093/mnras/stad1677}, \href {https://ui.adsabs.harvard.edu/abs/2023MNRAS.524..886H} {524, 886}

\bibitem[\protect\citeauthoryear{{Holden}, {Tadhunter}, {Morganti}  \& {Oosterloo}}{{Holden} et~al.}{2023}]{holden2023b}
{Holden} L.~R.,  {Tadhunter} C.~N.,  {Morganti} R.,   {Oosterloo} T.,  2023, \mn@doi [\mnras] {10.1093/mnras/stad123}, \href {https://ui.adsabs.harvard.edu/abs/2023MNRAS.520.1848H} {520, 1848}

\bibitem[\protect\citeauthoryear{{Hollenbach} \& {McKee}}{{Hollenbach} \& {McKee}}{1989}]{hollenbach1989}
{Hollenbach} D.,  {McKee} C.~F.,  1989, \mn@doi [\apj] {10.1086/167595}, \href {https://ui.adsabs.harvard.edu/abs/1989ApJ...342..306H} {342, 306}

\bibitem[\protect\citeauthoryear{{Hopkins} \& {Elvis}}{{Hopkins} \& {Elvis}}{2010}]{hopkins2010}
{Hopkins} P.~F.,  {Elvis} M.,  2010, \mn@doi [\mnras] {10.1111/j.1365-2966.2009.15643.x}, \href {https://ui.adsabs.harvard.edu/abs/2010MNRAS.401....7H} {401, 7}

\bibitem[\protect\citeauthoryear{{Ichikawa}, {Ricci}, {Ueda}, {Matsuoka}, {Toba}, {Kawamuro}, {Trakhtenbrot}  \& {Koss}}{{Ichikawa} et~al.}{2017}]{ichikawa2017}
{Ichikawa} K.,  {Ricci} C.,  {Ueda} Y.,  {Matsuoka} K.,  {Toba} Y.,  {Kawamuro} T.,  {Trakhtenbrot} B.,   {Koss} M.~J.,  2017, \mn@doi [\apj] {10.3847/1538-4357/835/1/74}, \href {https://ui.adsabs.harvard.edu/abs/2017ApJ...835...74I} {835, 74}

\bibitem[\protect\citeauthoryear{{Jarrett}, {Chester}, {Cutri}, {Schneider}  \& {Huchra}}{{Jarrett} et~al.}{2003}]{jarret03}
{Jarrett} T.~H.,  {Chester} T.,  {Cutri} R.,  {Schneider} S.~E.,   {Huchra} J.~P.,  2003, \mn@doi [\aj] {10.1086/345794}, \href {https://ui.adsabs.harvard.edu/abs/2003AJ....125..525J} {125, 525}

\bibitem[\protect\citeauthoryear{{Kakkad} et~al.,}{{Kakkad} et~al.}{2018}]{kakkad18}
{Kakkad} D.,  et~al., 2018, \mn@doi [\aap] {10.1051/0004-6361/201832790}, \href {https://ui.adsabs.harvard.edu/abs/2018A&A...618A...6K} {618, A6}

\bibitem[\protect\citeauthoryear{{Kakkad} et~al.,}{{Kakkad} et~al.}{2020}]{kakkad2020}
{Kakkad} D.,  et~al., 2020, \mn@doi [\aap] {10.1051/0004-6361/202038551}, \href {https://ui.adsabs.harvard.edu/abs/2020A&A...642A.147K} {642, A147}

\bibitem[\protect\citeauthoryear{{Kormendy} \& {Ho}}{{Kormendy} \& {Ho}}{2013}]{kormendy2013}
{Kormendy} J.,  {Ho} L.~C.,  2013, \mn@doi [\araa] {10.1146/annurev-astro-082708-101811}, \href {https://ui.adsabs.harvard.edu/abs/2013ARA&A..51..511K} {51, 511}

\bibitem[\protect\citeauthoryear{{Larkin}, {Armus}, {Knop}, {Soifer}  \& {Matthews}}{{Larkin} et~al.}{1998}]{larkin1998}
{Larkin} J.~E.,  {Armus} L.,  {Knop} R.~A.,  {Soifer} B.~T.,   {Matthews} K.,  1998, \mn@doi [\apjs] {10.1086/313063}, \href {https://ui.adsabs.harvard.edu/abs/1998ApJS..114...59L} {114, 59}

\bibitem[\protect\citeauthoryear{{Liu}, {Zakamska}, {Greene}, {Nesvadba}  \& {Liu}}{{Liu} et~al.}{2013}]{liu2013}
{Liu} G.,  {Zakamska} N.~L.,  {Greene} J.~E.,  {Nesvadba} N. P.~H.,   {Liu} X.,  2013, \mn@doi [\mnras] {10.1093/mnras/stt1755}, \href {https://ui.adsabs.harvard.edu/abs/2013MNRAS.436.2576L} {436, 2576}

\bibitem[\protect\citeauthoryear{{Lutz} et~al.,}{{Lutz} et~al.}{2020}]{lutz2020}
{Lutz} D.,  et~al., 2020, \mn@doi [\aap] {10.1051/0004-6361/201936803}, \href {https://ui.adsabs.harvard.edu/abs/2020A&A...633A.134L} {633, A134}

\bibitem[\protect\citeauthoryear{{Maiolino} et~al.,}{{Maiolino} et~al.}{2017}]{maiolino2017}
{Maiolino} R.,  et~al., 2017, \mn@doi [\nat] {10.1038/nature21677}, \href {https://ui.adsabs.harvard.edu/abs/2017Natur.544..202M} {544, 202}

\bibitem[\protect\citeauthoryear{{Maloney}, {Hollenbach}  \& {Tielens}}{{Maloney} et~al.}{1996}]{maloney1996}
{Maloney} P.~R.,  {Hollenbach} D.~J.,   {Tielens} A.~G.~G.~M.,  1996, \mn@doi [\apj] {10.1086/177532}, \href {https://ui.adsabs.harvard.edu/abs/1996ApJ...466..561M} {466, 561}

\bibitem[\protect\citeauthoryear{{Mazzalay} et~al.,}{{Mazzalay} et~al.}{2014}]{mazzalay2014}
{Mazzalay} X.,  et~al., 2014, \mn@doi [\mnras] {10.1093/mnras/stt2319}, \href {https://ui.adsabs.harvard.edu/abs/2014MNRAS.438.2036M} {438, 2036}

\bibitem[\protect\citeauthoryear{{McGregor} et~al.,}{{McGregor} et~al.}{2003}]{mcgregor2003}
{McGregor} P.~J.,  et~al., 2003, in {Iye} M.,  {Moorwood} A. F.~M.,  eds,  Society of Photo-Optical Instrumentation Engineers (SPIE) Conference Series Vol. 4841, Instrument Design and Performance for Optical/Infrared Ground-based Telescopes. pp 1581--1591, \mn@doi{10.1117/12.459448}

\bibitem[\protect\citeauthoryear{{Mouri}, {Nishida}, {Taniguchi}  \& {Kawara}}{{Mouri} et~al.}{1990}]{mouri1990}
{Mouri} H.,  {Nishida} M.,  {Taniguchi} Y.,   {Kawara} K.,  1990, \mn@doi [\apj] {10.1086/169095}, \href {https://ui.adsabs.harvard.edu/abs/1990ApJ...360...55M} {360, 55}

\bibitem[\protect\citeauthoryear{{Mouri}, {Kawara}  \& {Taniguchi}}{{Mouri} et~al.}{1993}]{mouri1993}
{Mouri} H.,  {Kawara} K.,   {Taniguchi} Y.,  1993, \mn@doi [\apj] {10.1086/172419}, \href {https://ui.adsabs.harvard.edu/abs/1993ApJ...406...52M} {406, 52}

\bibitem[\protect\citeauthoryear{{Mulchaey}, {Wilson}  \& {Tsvetanov}}{{Mulchaey} et~al.}{1996}]{mulchaey1996}
{Mulchaey} J.~S.,  {Wilson} A.~S.,   {Tsvetanov} Z.,  1996, \mn@doi [\apjs] {10.1086/192261}, \href {https://ui.adsabs.harvard.edu/abs/1996ApJS..102..309M} {102, 309}

\bibitem[\protect\citeauthoryear{{Nandi} et~al.,}{{Nandi} et~al.}{2023a}]{Nandi23}
{Nandi} P.,  et~al., 2023a, \mn@doi [\apj] {10.3847/1538-4357/ad0c57}, \href {https://ui.adsabs.harvard.edu/abs/2023ApJ...959..116N} {959, 116}

\bibitem[\protect\citeauthoryear{{Nandi} et~al.,}{{Nandi} et~al.}{2023b}]{nandi2023}
{Nandi} P.,  et~al., 2023b, \mn@doi [\apj] {10.3847/1538-4357/ad0c57}, \href {https://ui.adsabs.harvard.edu/abs/2023ApJ...959..116N} {959, 116}

\bibitem[\protect\citeauthoryear{{Oh} et~al.,}{{Oh} et~al.}{2018}]{oh2018}
{Oh} K.,  et~al., 2018, \mn@doi [\apjs] {10.3847/1538-4365/aaa7fd}, \href {https://ui.adsabs.harvard.edu/abs/2018ApJS..235....4O} {235, 4}

\bibitem[\protect\citeauthoryear{{Osterbrock} \& {Ferland}}{{Osterbrock} \& {Ferland}}{2006}]{osterbrock2006}
{Osterbrock} D.~E.,  {Ferland} G.~J.,  2006, {Astrophysics of gaseous nebulae and active galactic nuclei}

\bibitem[\protect\citeauthoryear{{Reunanen}, {Kotilainen}  \& {Prieto}}{{Reunanen} et~al.}{2002}]{reunanen2002}
{Reunanen} J.,  {Kotilainen} J.~K.,   {Prieto} M.~A.,  2002, \mn@doi [\mnras] {10.1046/j.1365-8711.2002.05181.x}, \href {https://ui.adsabs.harvard.edu/abs/2002MNRAS.331..154R} {331, 154}

\bibitem[\protect\citeauthoryear{Revalski et~al.,}{Revalski et~al.}{2022}]{revalski2022}
Revalski M.,  et~al., 2022, \mn@doi [The Astrophysical Journal] {10.3847/1538-4357/ac5f3d}, 930, 14

\bibitem[\protect\citeauthoryear{{Richings} \& {Faucher-Gigu{\`e}re}}{{Richings} \& {Faucher-Gigu{\`e}re}}{2018}]{richings2018}
{Richings} A.~J.,  {Faucher-Gigu{\`e}re} C.-A.,  2018, \mn@doi [\mnras] {10.1093/mnras/sty1285}, \href {https://ui.adsabs.harvard.edu/abs/2018MNRAS.478.3100R} {478, 3100}

\bibitem[\protect\citeauthoryear{Riffel \& Storchi-Bergmann}{Riffel \& Storchi-Bergmann}{2011}]{riffel2011}
Riffel R.~A.,  Storchi-Bergmann T.,  2011, \mn@doi [Monthly Notices of the Royal Astronomical Society] {10.1111/j.1365-2966.2010.17721.x}, 411, 469

\bibitem[\protect\citeauthoryear{{Riffel, R.}, {Rodríguez-Ardila, A.}  \& {Pastoriza, M. G.}}{{Riffel, R.} et~al.}{2006}]{rogerio2006}
{Riffel, R.} {Rodríguez-Ardila, A.}  {Pastoriza, M. G.} 2006, \mn@doi [A&A] {10.1051/0004-6361:20065291}, 457, 61

\bibitem[\protect\citeauthoryear{{Riffel}, {Storchi-Bergmann}, {Winge}  \& {Barbosa}}{{Riffel} et~al.}{2006}]{riffel2006}
{Riffel} R.~A.,  {Storchi-Bergmann} T.,  {Winge} C.,   {Barbosa} F. K.~B.,  2006, \mn@doi [\mnras] {10.1111/j.1365-2966.2006.11050.x}, \href {https://ui.adsabs.harvard.edu/abs/2006MNRAS.373....2R} {373, 2}

\bibitem[\protect\citeauthoryear{{Riffel}, {Storchi-Bergmann}, {Winge}, {McGregor}, {Beck}  \& {Schmitt}}{{Riffel} et~al.}{2008}]{riffel2008}
{Riffel} R.~A.,  {Storchi-Bergmann} T.,  {Winge} C.,  {McGregor} P.~J.,  {Beck} T.,   {Schmitt} H.,  2008, \mn@doi [\mnras] {10.1111/j.1365-2966.2008.12936.x}, \href {https://ui.adsabs.harvard.edu/abs/2008MNRAS.385.1129R} {385, 1129}

\bibitem[\protect\citeauthoryear{{Riffel}, {Storchi-Bergmann}  \& {Nagar}}{{Riffel} et~al.}{2010}]{riffel2010}
{Riffel} R.~A.,  {Storchi-Bergmann} T.,   {Nagar} N.~M.,  2010, \mn@doi [\mnras] {10.1111/j.1365-2966.2010.16308.x}, \href {https://ui.adsabs.harvard.edu/abs/2010MNRAS.404..166R} {404, 166}

\bibitem[\protect\citeauthoryear{{Riffel}, {Rodr{\'\i}guez-Ardila}, {Aleman}, {Brotherton}, {Pastoriza}, {Bonatto}  \& {Dors}}{{Riffel} et~al.}{2013b}]{riffel2013a}
{Riffel} R.,  {Rodr{\'\i}guez-Ardila} A.,  {Aleman} I.,  {Brotherton} M.~S.,  {Pastoriza} M.~G.,  {Bonatto} C.,   {Dors} O.~L.,  2013b, \mn@doi [\mnras] {10.1093/mnras/stt026}, \href {https://ui.adsabs.harvard.edu/abs/2013MNRAS.430.2002R} {430, 2002}

\bibitem[\protect\citeauthoryear{Riffel, Rodríguez-Ardila, Aleman, Brotherton, Pastoriza, Bonatto  \& Dors}{Riffel et~al.}{2013a}]{rogerio2013}
Riffel R.,  Rodríguez-Ardila A.,  Aleman I.,  Brotherton M.~S.,  Pastoriza M.~G.,  Bonatto C.,   Dors O.~L. J.,  2013a, \mn@doi [Monthly Notices of the Royal Astronomical Society] {10.1093/mnras/stt026}, 430, 2002

\bibitem[\protect\citeauthoryear{{Riffel}, {Storchi-Bergmann}  \& {Winge}}{{Riffel} et~al.}{2013c}]{riffel2013}
{Riffel} R.~A.,  {Storchi-Bergmann} T.,   {Winge} C.,  2013c, \mn@doi [\mnras] {10.1093/mnras/stt045}, \href {https://ui.adsabs.harvard.edu/abs/2013MNRAS.430.2249R} {430, 2249}

\bibitem[\protect\citeauthoryear{Riffel, Vale, Storchi-Bergmann  \& McGregor}{Riffel et~al.}{2014}]{riffel2014}
Riffel R.~A.,  Vale T.~B.,  Storchi-Bergmann T.,   McGregor P.~J.,  2014, \mn@doi [Monthly Notices of the Royal Astronomical Society] {10.1093/mnras/stu843}, 442, 656

\bibitem[\protect\citeauthoryear{Riffel, Storchi-Bergmann, Riffel, Dahmer-Hahn, Diniz, Schönell  \& Dametto}{Riffel et~al.}{2017}]{riffel2017}
Riffel R.~A.,  Storchi-Bergmann T.,  Riffel R.,  Dahmer-Hahn L.~G.,  Diniz M.~R.,  Schönell A.~J.,   Dametto N.~Z.,  2017, \mn@doi [Monthly Notices of the Royal Astronomical Society] {10.1093/mnras/stx1308}, 470, 992

\bibitem[\protect\citeauthoryear{{Riffel} et~al.,}{{Riffel} et~al.}{2018}]{riffel2018}
{Riffel} R.~A.,  et~al., 2018, \mn@doi [\mnras] {10.1093/mnras/stx2857}, \href {https://ui.adsabs.harvard.edu/abs/2018MNRAS.474.1373R} {474, 1373}

\bibitem[\protect\citeauthoryear{{Riffel}, {Storchi-Bergmann}, {Zakamska}  \& {Riffel}}{{Riffel} et~al.}{2020}]{riffel2020}
{Riffel} R.~A.,  {Storchi-Bergmann} T.,  {Zakamska} N.~L.,   {Riffel} R.,  2020, \mn@doi [\mnras] {10.1093/mnras/staa1922}, \href {https://ui.adsabs.harvard.edu/abs/2020MNRAS.496.4857R} {496, 4857}

\bibitem[\protect\citeauthoryear{{Riffel}, {Bianchin}, {Riffel}, {Storchi-Bergmann}, {Sch{\"o}nell}, {Dahmer-Hahn}, {Dametto}  \& {Diniz}}{{Riffel} et~al.}{2021b}]{riffel2021}
{Riffel} R.~A.,  {Bianchin} M.,  {Riffel} R.,  {Storchi-Bergmann} T.,  {Sch{\"o}nell} A.~J.,  {Dahmer-Hahn} L.~G.,  {Dametto} N.~Z.,   {Diniz} M.~R.,  2021b, \mn@doi [\mnras] {10.1093/mnras/stab788}, \href {https://ui.adsabs.harvard.edu/abs/2021MNRAS.503.5161R} {503, 5161}

\bibitem[\protect\citeauthoryear{Riffel, Bianchin, Riffel, Storchi-Bergmann, Schönell, Dahmer-Hahn, Dametto  \& Diniz}{Riffel et~al.}{2021a}]{riffel21}
Riffel R.~A.,  Bianchin M.,  Riffel R.,  Storchi-Bergmann T.,  Schönell A.~J.,  Dahmer-Hahn L.~G.,  Dametto N.~Z.,   Diniz M.~R.,  2021a, \mn@doi [Monthly Notices of the Royal Astronomical Society] {10.1093/mnras/stab788}, 503, 5161

\bibitem[\protect\citeauthoryear{Riffel et~al.,}{Riffel et~al.}{2021c}]{riffel21b}
Riffel R.~A.,  et~al., 2021c, \mn@doi [Monthly Notices of the Royal Astronomical Society] {10.1093/mnras/stab998}, 504, 3265

\bibitem[\protect\citeauthoryear{Riffel et~al.,}{Riffel et~al.}{2022}]{rogerio22}
Riffel R.,  et~al., 2022, \mn@doi [Monthly Notices of the Royal Astronomical Society] {10.1093/mnras/stac740}, 512, 3906

\bibitem[\protect\citeauthoryear{{Riffel} et~al.,}{{Riffel} et~al.}{2023}]{riffel2023}
{Riffel} R.~A.,  et~al., 2023, \mn@doi [\mnras] {10.1093/mnras/stad599}, \href {https://ui.adsabs.harvard.edu/abs/2023MNRAS.521.1832R} {521, 1832}

\bibitem[\protect\citeauthoryear{{Rodr{\'\i}guez-Ardila}, {Pastoriza}, {Viegas}, {Sigut}  \& {Pradhan}}{{Rodr{\'\i}guez-Ardila} et~al.}{2004}]{ardila2004}
{Rodr{\'\i}guez-Ardila} A.,  {Pastoriza} M.~G.,  {Viegas} S.,  {Sigut} T.~A.~A.,   {Pradhan} A.~K.,  2004, \mn@doi [\aap] {10.1051/0004-6361:20034285}, \href {https://ui.adsabs.harvard.edu/abs/2004A&A...425..457R} {425, 457}

\bibitem[\protect\citeauthoryear{{Rodr{\'\i}guez-Ardila}, {Riffel}  \& {Pastoriza}}{{Rodr{\'\i}guez-Ardila} et~al.}{2005}]{ardila2005}
{Rodr{\'\i}guez-Ardila} A.,  {Riffel} R.,   {Pastoriza} M.~G.,  2005, \mn@doi [\mnras] {10.1111/j.1365-2966.2005.09638.x}, \href {https://ui.adsabs.harvard.edu/abs/2005MNRAS.364.1041R} {364, 1041}

\bibitem[\protect\citeauthoryear{Rodríguez-Ardila et~al.,}{Rodríguez-Ardila et~al.}{2016}]{ardila2016}
Rodríguez-Ardila A.,  et~al., 2016, \mn@doi [Monthly Notices of the Royal Astronomical Society] {10.1093/mnras/stw2642}, 465, 906

\bibitem[\protect\citeauthoryear{{Rose}, {Tadhunter}, {Ramos Almeida}, {Rodr{\'\i}guez Zaur{\'\i}n}, {Santoro}  \& {Spence}}{{Rose} et~al.}{2018}]{rose2018}
{Rose} M.,  {Tadhunter} C.,  {Ramos Almeida} C.,  {Rodr{\'\i}guez Zaur{\'\i}n} J.,  {Santoro} F.,   {Spence} R.,  2018, \mn@doi [\mnras] {10.1093/mnras/stx2590}, \href {https://ui.adsabs.harvard.edu/abs/2018MNRAS.474..128R} {474, 128}

\bibitem[\protect\citeauthoryear{Ruschel-Dutra et~al.,}{Ruschel-Dutra et~al.}{2021}]{rd2021}
Ruschel-Dutra D.,  et~al., 2021, \mn@doi [Monthly Notices of the Royal Astronomical Society] {10.1093/mnras/stab2058}, 507, 74

\bibitem[\protect\citeauthoryear{{Santoro}, {Tadhunter}, {Baron}, {Morganti}  \& {Holt}}{{Santoro} et~al.}{2020}]{santoro2020}
{Santoro} F.,  {Tadhunter} C.,  {Baron} D.,  {Morganti} R.,   {Holt} J.,  2020, \mn@doi [\aap] {10.1051/0004-6361/202039077}, \href {https://ui.adsabs.harvard.edu/abs/2020A&A...644A..54S} {644, A54}

\bibitem[\protect\citeauthoryear{{Schaye} et~al.,}{{Schaye} et~al.}{2015}]{schaye2015}
{Schaye} J.,  et~al., 2015, \mn@doi [\mnras] {10.1093/mnras/stu2058}, \href {https://ui.adsabs.harvard.edu/abs/2015MNRAS.446..521S} {446, 521}

\bibitem[\protect\citeauthoryear{{Sch{\"o}nell}, {Riffel}, {Storchi-Bergmann}  \& {Winge}}{{Sch{\"o}nell} et~al.}{2014}]{schonell2014}
{Sch{\"o}nell} A.~J.,  {Riffel} R.~A.,  {Storchi-Bergmann} T.,   {Winge} C.,  2014, \mn@doi [\mnras] {10.1093/mnras/stu1685}, \href {https://ui.adsabs.harvard.edu/abs/2014MNRAS.445..414S} {445, 414}

\bibitem[\protect\citeauthoryear{{Sch{\"o}nell}, {Storchi-Bergmann}, {Riffel}, {Riffel}, {Bianchin}, {Dahmer-Hahn}, {Diniz}  \& {Dametto}}{{Sch{\"o}nell} et~al.}{2019}]{schonell2019}
{Sch{\"o}nell} A.~J.,  {Storchi-Bergmann} T.,  {Riffel} R.~A.,  {Riffel} R.,  {Bianchin} M.,  {Dahmer-Hahn} L.~G.,  {Diniz} M.~R.,   {Dametto} N.~Z.,  2019, \mn@doi [\mnras] {10.1093/mnras/stz523}, \href {https://ui.adsabs.harvard.edu/abs/2019MNRAS.485.2054S} {485, 2054}

\bibitem[\protect\citeauthoryear{Schönell, Storchi-Bergmann, Riffel  \& Riffel}{Schönell et~al.}{2016}]{schonell2017}
Schönell Astor~J. J.,  Storchi-Bergmann T.,  Riffel R.~A.,   Riffel R.,  2016, \mn@doi [Monthly Notices of the Royal Astronomical Society] {10.1093/mnras/stw2263}, 464, 1771

\bibitem[\protect\citeauthoryear{{Silk} \& {Mamon}}{{Silk} \& {Mamon}}{2012}]{silk2012}
{Silk} J.,  {Mamon} G.~A.,  2012, \mn@doi [Research in Astronomy and Astrophysics] {10.1088/1674-4527/12/8/004}, \href {https://ui.adsabs.harvard.edu/abs/2012RAA....12..917S} {12, 917}

\bibitem[\protect\citeauthoryear{{Simpson}, {Forbes}, {Baker}  \& {Ward}}{{Simpson} et~al.}{1996}]{simpson1996}
{Simpson} C.,  {Forbes} D.~A.,  {Baker} A.~C.,   {Ward} M.~J.,  1996, \mn@doi [\mnras] {10.1093/mnras/283.3.777}, \href {https://ui.adsabs.harvard.edu/abs/1996MNRAS.283..777S} {283, 777}

\bibitem[\protect\citeauthoryear{{Somerville}, {Hopkins}, {Cox}, {Robertson}  \& {Hernquist}}{{Somerville} et~al.}{2008}]{somerville2008}
{Somerville} R.~S.,  {Hopkins} P.~F.,  {Cox} T.~J.,  {Robertson} B.~E.,   {Hernquist} L.,  2008, \mn@doi [\mnras] {10.1111/j.1365-2966.2008.13805.x}, \href {https://ui.adsabs.harvard.edu/abs/2008MNRAS.391..481S} {391, 481}

\bibitem[\protect\citeauthoryear{{Sternberg} \& {Dalgarno}}{{Sternberg} \& {Dalgarno}}{1989}]{sternberg1989}
{Sternberg} A.,  {Dalgarno} A.,  1989, \mn@doi [\apj] {10.1086/167193}, \href {https://ui.adsabs.harvard.edu/abs/1989ApJ...338..197S} {338, 197}

\bibitem[\protect\citeauthoryear{{Storchi-Bergmann}, {McGregor}, {Riffel}, {Sim{\~o}es Lopes}, {Beck}  \& {Dopita}}{{Storchi-Bergmann} et~al.}{2009a}]{storchi2009}
{Storchi-Bergmann} T.,  {McGregor} P.~J.,  {Riffel} R.~A.,  {Sim{\~o}es Lopes} R.,  {Beck} T.,   {Dopita} M.,  2009a, \mn@doi [\mnras] {10.1111/j.1365-2966.2009.14388.x}, \href {https://ui.adsabs.harvard.edu/abs/2009MNRAS.394.1148S} {394, 1148}

\bibitem[\protect\citeauthoryear{{Storchi-Bergmann}, {McGregor}, {Riffel}, {Sim{\~o}es Lopes}, {Beck}  \& {Dopita}}{{Storchi-Bergmann} et~al.}{2009b}]{sb2009}
{Storchi-Bergmann} T.,  {McGregor} P.~J.,  {Riffel} R.~A.,  {Sim{\~o}es Lopes} R.,  {Beck} T.,   {Dopita} M.,  2009b, \mn@doi [\mnras] {10.1111/j.1365-2966.2009.14388.x}, \href {https://ui.adsabs.harvard.edu/abs/2009MNRAS.394.1148S} {394, 1148}

\bibitem[\protect\citeauthoryear{{Storchi-Bergmann}, {Lopes}, {McGregor}, {Riffel}, {Beck}  \& {Martini}}{{Storchi-Bergmann} et~al.}{2010}]{sb2010}
{Storchi-Bergmann} T.,  {Lopes} R.~D.~S.,  {McGregor} P.~J.,  {Riffel} R.~A.,  {Beck} T.,   {Martini} P.,  2010, \mn@doi [\mnras] {10.1111/j.1365-2966.2009.15962.x}, \href {https://ui.adsabs.harvard.edu/abs/2010MNRAS.402..819S} {402, 819}

\bibitem[\protect\citeauthoryear{{Thean}, {Pedlar}, {Kukula}, {Baum}  \& {O'Dea}}{{Thean} et~al.}{2000}]{thean2000}
{Thean} A.,  {Pedlar} A.,  {Kukula} M.~J.,  {Baum} S.~A.,   {O'Dea} C.~P.,  2000, \mn@doi [\mnras] {10.1046/j.1365-8711.2000.03401.x}, \href {https://ui.adsabs.harvard.edu/abs/2000MNRAS.314..573T} {314, 573}

\bibitem[\protect\citeauthoryear{{Theureau}, {Bottinelli}, {Coudreau-Durand}, {Gouguenheim}, {Hallet}, {Loulergue}, {Paturel}  \& {Teerikorpi}}{{Theureau} et~al.}{1998}]{theureau98}
{Theureau} G.,  {Bottinelli} L.,  {Coudreau-Durand} N.,  {Gouguenheim} L.,  {Hallet} N.,  {Loulergue} M.,  {Paturel} G.,   {Teerikorpi} P.,  1998, \mn@doi [\aaps] {10.1051/aas:1998416}, \href {https://ui.adsabs.harvard.edu/abs/1998A&AS..130..333T} {130, 333}

\bibitem[\protect\citeauthoryear{{Tody}}{{Tody}}{1986}]{tody86}
{Tody} D.,  1986, in {Crawford} D.~L.,  ed.,  Society of Photo-Optical Instrumentation Engineers (SPIE) Conference Series Vol. 627, Instrumentation in astronomy VI. p.~733, \mn@doi{10.1117/12.968154}

\bibitem[\protect\citeauthoryear{{Tody}}{{Tody}}{1993}]{tody93}
{Tody} D.,  1993, in {Hanisch} R.~J.,  {Brissenden} R.~J.~V.,   {Barnes} J.,  eds,  Astronomical Society of the Pacific Conference Series Vol. 52, Astronomical Data Analysis Software and Systems II. p.~173

\bibitem[\protect\citeauthoryear{{Weinberger} et~al.,}{{Weinberger} et~al.}{2017}]{weinberger2017}
{Weinberger} R.,  et~al., 2017, \mn@doi [\mnras] {10.1093/mnras/stw2944}, \href {https://ui.adsabs.harvard.edu/abs/2017MNRAS.465.3291W} {465, 3291}

\bibitem[\protect\citeauthoryear{{Winge}, {Riffel}  \& {Storchi-Bergmann}}{{Winge} et~al.}{2009}]{winge2009}
{Winge} C.,  {Riffel} R.~A.,   {Storchi-Bergmann} T.,  2009, \mn@doi [\apjs] {10.1088/0067-0049/185/1/186}, \href {https://ui.adsabs.harvard.edu/abs/2009ApJS..185..186W} {185, 186}

\bibitem[\protect\citeauthoryear{{van der Kruit} \& {Allen}}{{van der Kruit} \& {Allen}}{1978}]{vdk1978}
{van der Kruit} P.~C.,  {Allen} R.~J.,  1978, \mn@doi [\araa] {10.1146/annurev.aa.16.090178.000535}, \href {https://ui.adsabs.harvard.edu/abs/1978ARA&A..16..103V} {16, 103}

\makeatother
\end{thebibliography}

%\begin{thebibliography}{99}
%\end{thebibliography}

%%%%%%%%%%%%%%%%%%%%%%%%%%%%%%%%%%%%%%%%%%%%%%%%%%

%%%%%%%%%%%%%%%%% APPENDICES %%%%%%%%%%%%%%%%%%%%%

%\appendix


%if you want to present additional material which would interrupt the flow of the main paper,

%%%%%%%%%%%%%%%%%%%%%%%%%%%%%%%%%%%%%%%%%%%%%%%%%%

% Don't change these lines
\bsp	% typesetting comment
\label{lastpage}
\end{document}